\documentclass[preprint,12pt]{elsarticle}

\usepackage{amsmath,amssymb}
\usepackage{bm}
\usepackage{url}

\usepackage{color}

\newcommand{\RevisionA}[1]{\textcolor{black}{#1}}
\newcommand{\RevisionB}[1]{\textcolor{black}{#1}}

\usepackage{setspace}
\journal{Building and Environment}

\begin{document}

\begin{frontmatter}



\title{\RevisionA{Super-Resolution of Near-Surface Temperature Utilizing Physical Quantities for Real-Time Prediction of Urban Micrometeorology}}


\affiliation[inst1]{
    organization={Global Scientific Information and Computing Center, Tokyo Institute of Technology},
    addressline={2-12-1 Ookayama, Meguro-ku},
    city={Tokyo},
    postcode={1528550},
    country={Japan}
}

\affiliation[inst2]{
    organization={Ashikaga University},
    addressline={268-1 Omae-cho, Ashikaga-shi},
    city={Tochigi},
    postcode={3268558},
    country={Japan}
}

\affiliation[inst3]{
    organization={Center for Design, Manufacturing and Materials, Skolkovo Institute of Science and Technology},
    addressline={Bld. 1, 30 Bolshoy Boulevard},
    city={Moscow},
    postcode={121205},
    country={Russia}
}

\affiliation[inst4]{
    organization={Research Institute of Value-Added Information Generation, Japan Agency for Marine-Earth Science and Technology},
    addressline={3173-25 Showa-machi, Kanazawa-ku, Yokohama},
    city={Kanagawa},
    postcode={2360001},
    country={Japan}
}

\author[inst1]{Yuki Yasuda\fnref{corresponding_authoer_info}}
\ead{yasuda.y.aa@m.titech.ac.jp}
\fntext[Corresponding_authoer_info]{Corresponding author. 2-12-1-i9-111 Ookayama, Meguro-ku, Tokyo, 1528550, Japan. Tel.: +81 3 57343173 (office). Email address: yasuda.y.aa@m.titech.ac.jp}

\author[inst1]{Ryo Onishi}
\ead{onishi.ryo@gsic.titech.ac.jp}

\author[inst2]{Yuichi Hirokawa}
\ead{hirokawa.yuichi@g.ashikaga.ac.jp}

\author[inst3]{Dmitry Kolomenskiy}
\ead{d.kolomenskiy@skoltech.ru}

\author[inst4]{Daisuke Sugiyama}
\ead{sugiyamad@jamstec.go.jp}

\begin{abstract}
% \begin{linenumbers}
The present paper proposes a super-resolution (SR) model based on a convolutional neural network and applies it to the near-surface temperature in urban areas. The SR model incorporates a skip connection, a channel attention mechanism, and separated feature extractors for the inputs of temperature, building height, downward shortwave radiation, and horizontal velocity. We train the SR model with sets of low-resolution (LR) and high-resolution (HR) images from building-resolving large-eddy simulations (LESs) in a city, \RevisionA{where the horizontal resolutions of LR and HR are 20 and 5 m, respectively.} The generalization capability of the SR model is confirmed with LESs in another city. The estimated HR temperature fields are more accurate than those of the bicubic interpolation and image SR model that takes only the temperature as its input. Except for the temperature input, the building height is the most important to reconstruct the HR temperature and enables the SR model to reduce errors in temperature near building boundaries. \RevisionA{The SR model considers the appropriate boundary for each building from its height information.} The analysis of attention weights indicates that the importance of the building height increases as the downward shortwave radiation becomes larger. The contrast between sun and shade is strengthened with the increase in solar radiation, which may affect the temperature distribution. The short inference time suggests the potential of the proposed SR model to facilitate a real-time HR prediction in metropolitan areas by combining it with an LR building-resolving LES model.
% \end{linenumbers}
\end{abstract}


% \begin{highlights}
% \end{highlights}

\begin{keyword}
%% keywords here, in the form: keyword \sep keyword (Max of keywords is 6)
super-resolution \sep downscaling \sep building-resolving micrometeorological model \sep large-eddy simulation \sep artificial neural network \sep attention mechanism
% \PACS 47.27.E- \sep 92.90.+x
\end{keyword}

\end{frontmatter}


\section{Introduction} \label{sec:intrudoction}

Urban built infrastructure has numerous important impacts on the environment and consequently on human well-being. The impacts include changes in the micrometeorological conditions such as urban wind and heat island effects. In particular, the urban heat island effect in combination with the global warming has led to heat stress becoming a rapidly growing social problem. Nowadays, architects and urban planners address the building design taking into account the knowledge of urban micrometeorology, including the thermal and dynamical airflow response of the urban environment \citep{Qu2011, Naboni2013, Toparlar2017}. Among the variety of tools employed, numerical simulation approaches based on computational fluid dynamics (CFD) play an important role. However, despite the ever-increasing computer performance, building-resolving CFD simulations require computational resources largely beyond the capabilities of desktop workstations. Modest computational resources can produce only low-resolution prediction maps with spatial grid steps of no less than several hundred meters. 

\RevisionA{Super-resolution (SR) refers to a method of estimating high-resolution images from low-resolution ones and has been actively studied in the field of computer vision in recent years as an application of artificial neural networks (ANNs) \citep{Dong2014, Dong2016FSRCNN, Kim2016VDSR, Kim2016DRCN, Ledig2017SRGAN, Lim2017EDSR, Zhang2018RDN, Zhang2018RCAN, Wang2018ESRGAN, Dai2019SAN, Zhang2019RNAN, Chen2021A2N}.} The success of ANN-based SR methods has led to an increasing number of studies applying these SR methods to fluid-related problems: idealized turbulent flows in two \citep{Deng2019, Fukami2019, Maulik2020, Wang2020, Fukami2020, Fukami2021} and three dimensions \citep{Fukami2020, Fukami2021, Liu2020, Bode2021, Kim2021}, Rayleigh–B\'{e}nard convection \citep{Jiang2020}, smoke motions in turbulent flows \citep{Xie2018, Werhahn2019, Bai2020}, flows in blood vessels \citep{Ferdian2020, Sun2020, Gao2021}, sea surface temperature \citep{Ducournau2016, Maulik2020}, and atmospheric fields \RevisionA{\citep{Cannon2011, Vandal2017, Rodrigues2018, Onishi2019, Leinonen2020, Stengel2020}}. \RevisionA{Note that applications of machine learning, including SR, to fluids are discussed in comprehensive reviews \citep{Brunton2020, Duraisamy2021}. These SR methods may effectively infer high-resolution (HR) flow fields from low-resolution (LR) CFD simulations, reducing the computational cost.}

\RevisionA{We briefly review some recent studies on the ANN-based super-resolution for fluids.} Bode et al. \citep{Bode2021} developed a model based on a generative adversarial network (GAN) \citep{Wang2018ESRGAN} to super-resolve three-dimensional turbulent flows, in which the continuity equation was included in the loss function. Fukami et al. \citep{Fukami2021} applied their neural network \RevisionA{\citep{Fukami2019,Fukami2020}} sequentially in time and space and demonstrated that the spatio-temporal SR was possible for \RevisionA{three-dimensional channel} turbulent fields. \RevisionA{Kim et al. \citep{Kim2021} presented a cycle-consistent GAN and applied it to turbulent channel flows. They trained their model with unpaired HR and LR data and validated the super-resolved flow fields for a wide range of Reynolds numbers.} Gao et al. \citep{Gao2021} performed the super-resolution of stationary turbulence by incorporating the governing equations and boundary conditions into the loss function, without using HR data. These studies proposed state-of-the-art physics SR models, whereas these models were applied to relatively idealized flow fields.

ANN-based SR methods have also been developed for the atmosphere. For synoptic-scale precipitation, ANN-based models \citep{Vandal2017, Rodrigues2018, Leinonen2020} have inferred HR distributions with more precision than conventional downscaling methods. Leinonen et al. \citep{Leinonen2020} proposed a model based on a GAN and gated recurrent units \citep{Cho2014} to super-solve a time sequence of atmospheric fields. Their model was separately applied to precipitation and cloud optical thickness and generated HR fields that were spatially and temporally consistent with observations. Stengel et al. \citep{Stengel2020} applied a GAN-based model separately to the global wind velocity and solar irradiance, \RevisionA{and successfully increased resolution by a factor of 50 against the wind and 25 against the solar irradiance.} There are few SR studies in urban micrometeorology, and the effectiveness of super-resolution in this research area is not clear.

Although not directly related to meteorological problems, some researchers in remote sensing focus on the super-resolution of land surface temperature in urban areas. Xu et al. \citep{Xu2020} proposed an SR method combining a random forest \citep{Breiman2001} and Gaussian process regression \citep{Kyriakidis2004}, where the super-resolved temperature from the random forest was corrected with the Gaussian process regressor. They demonstrated that the super-resolution of 90 to 10 m was possible for the surface temperature over the targeted city. Yao et al. \citep{Yao2020} developed a monitoring method of surface urban heat islands by utilizing a random forest and some spatio-temporal data fusion approach \citep{Zhu2016}. They showed an increasing trend of the summer-daytime heat island intensity on the targeted city. Both studies investigated the super-resolution of observed land surface temperature and are distinguished from this paper utilizing micrometeorological simulation data.

\RevisionA{In our earlier study \citep{Onishi2019}, we investigated the super-resolution of near-surface temperature from building-resolving micrometeorological simulations, employing a convolutional neural network (CNN) \citep{Dong2014}. This CNN model reconstructed the near-surface temperature more accurately than algebraic interpolations such as the bicubic interpolation and Lanczos \citep{Duchon1979}. However, the input was temperature only and did not include other quantities such as velocity from the meteorological simulations; hence, there is room for improvement in the accuracy of super-resolution. We also proposed the following classification of SR methods: (Level 0) image SR method; (Level 1) physics SR that utilizes supplemental HR information; (Level 2) physics SR that utilizes statistical theories for mapping between LR and HR images; (Level 3) physics SR that directly utilizes governing equations. According to this classification, our previous CNN model falls into Level 0.}

The purpose of this paper is to confirm that the super-resolution with Level 1 (SR-lv1) estimates HR temperature fields more accurately than the image SR (SR-lv0) \RevisionA{in metropolitan areas by utilizing the dataset created by building-resolving micrometeorological simulations}. We improve over our previous image SR model \citep{Onishi2019} by incorporating a skip connection \citep{He2016} and a channel attention mechanism \citep{Hu2018}, as well as by adding more physical quantities as inputs such as HR building height and LR velocity. Our SR model was trained using the results of meteorological simulations in Tokyo and was evaluated with simulation results in another city to verify the generalization performance.

\RevisionA{The novelty of this research lies in demonstrating for the first time that a CNN-based model achieves a sufficiently accurate SR for near-surface temperatures obtained from real-world building-resolving micrometeorological simulations. In particular, we investigate in detail the effects of building height distributions on the accuracy of super-resolution.}

The rest of the paper is organized as follows. Section \ref{sec:sr_system_architecture} proposes our SR model. Section \ref{sec:training_data_preparation} explains configurations of micrometeorological simulations and gives a training method for the SR model. Section \ref{sec:results_discussion} presents the analysis of the results. Section \ref{sec:conclusions} summarizes the concluding remarks.

\section{Super-resolution system} \label{sec:sr_system_architecture}

\begin{figure}[htbp]
  \centering
   \includegraphics[width=14cm]{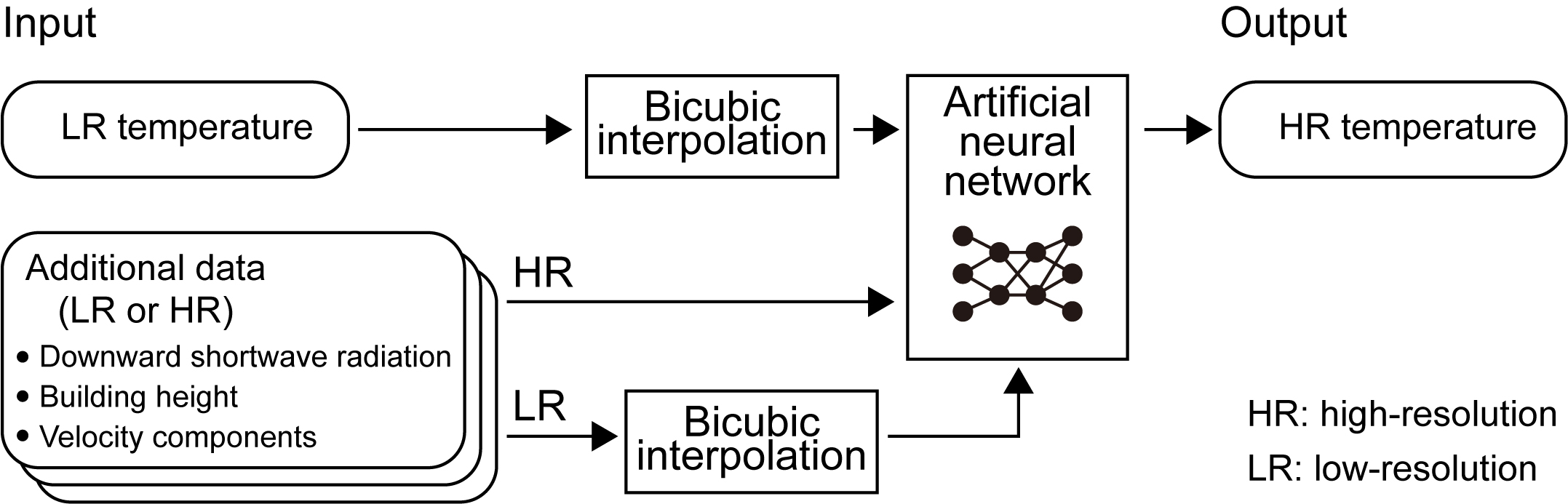}
  \caption{Data flows of the SR (super-resolution) system.}
  \label{fig:sr_system_data_flow}
\end{figure}

In this research, we estimate the temperature near the surface, which is one of the most important quantities for humans in metropolitan areas. Our SR system receives an LR temperature from a micrometeorological simulation and infers an HR temperature field by an ANN, as shown in Fig. \ref{fig:sr_system_data_flow}. As auxiliary input, we investigate effects of including the building height, the downward shortwave radiation at the surface, as well as the eastward and northward components of velocity near the surface, and combinations thereof. Fig. \ref{fig:sr_system_data_flow} shows that the ANN model receives some of these quantities. The building height and downward shortwave radiation are readily available in high-resolution, while the velocity components are in low-resolution and usually obtained jointly with the temperature in the same simulation. The LR and HR fields are encoded as monochrome PNG images with 256 color gradations. As a data preprocessing, all LR inputs are aligned in image size to the HR ones by bicubic interpolation. Following Onishi et al. \citep{Onishi2019}, we improve SRCNN (super-resolution CNN) \citep{Dong2014} as an ANN of the SR system (Fig. \ref{fig:sr_system_data_flow}), which is a regression-type SR model based on CNNs \citep{LeCun1989, Krizhevsky2012}. 

Improvements from the SRCNN in our previous study \citep{Onishi2019} are as follows:
\begin{itemize}
    \item adding a skip connection to learn residuals between LR and HR temperature fields.
    \item extracting features separately for each physical quantity.
    \item adjusting feature weights dynamically with a channel attention mechanism.
\end{itemize}

The skip connection \citep{He2016} has been widely used in both computer-vision \citep{Kim2016VDSR, Kim2016DRCN, Ledig2017SRGAN, Lim2017EDSR, Zhang2018RDN, Zhang2018RCAN, Wang2018ESRGAN, Dai2019SAN, Zhang2019RNAN, Chen2021A2N} and fluid-related SR \citep{Deng2019, Fukami2019, Wang2020, Fukami2020, Fukami2021, Liu2020, Bode2021, Jiang2020, Xie2018, Werhahn2019, Bai2020, Ferdian2020, Leinonen2020, Stengel2020}. Estimated HR and input LR images generally share similar large-scale signals. The skip connection can directly connect the input to the output, allowing SR models to learn the difference between the HR and LR images.

Our SR model receives different physical quantities in either high- or low-resolution. When receiving such different types of input, it may be natural to extract features separately. Indeed, this kind of separated branch in ANNs has sometimes been used in SR models \citep{Lim2017EDSR, Ferdian2020, Gao2021}. Moreover, this separation will facilitate transfer learning, where a pre-trained feature extractor can be included in an ANN for a similar SR task.

An SR model generally needs to handle more features (i.e., channels), after increasing the number of input physical quantities. The importance of these channels may vary depending on the input. It is desirable to infer an HR temperature by utilizing only the important channel information. Channel attention is a method dynamically weighting features over the channel direction. We apply a mechanism called squeeze-and-excitation (SE) block \citep{Hu2018}. It is worth noting that attention mechanisms are commonly used in the super-resolution of computer visions these days \citep{Zhang2018RCAN, Dai2019SAN, Zhang2019RNAN, Chen2021A2N} but rarely in fluid-related SR.

\RevisionA{The effectiveness of the skip connection and SE block is verified through an ablation study in Section \ref{subsec:ablation_study}. We show that the skip connection makes the training process more stable and faster, and the SE block improves the accuracy of super-resolution and reduces the number of outliers in inference. Furthermore, in Section \ref{subsec:discussion_attention}, we demonstrate that the importance of building height tends to increase as the downward shortwave radiation is stronger, which may be attributed to that the contrast between sun and shade made by buildings has a greater influence on the temperature distribution.}

\subsection{Super-resolution model architecture} \label{subsec:sr_model_architecture}

\begin{figure}[htbp]
  \centering
   \includegraphics[width=14cm]{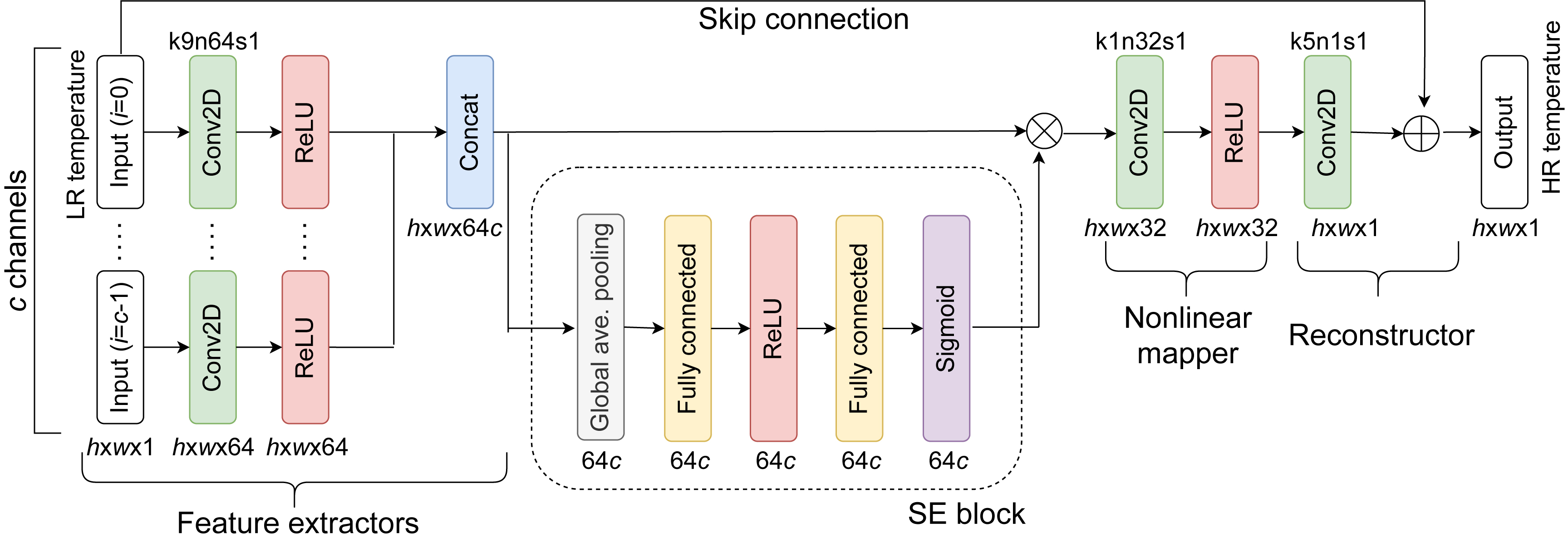}
  \caption{Architecture of the super-resolution model, namely squeeze-and-excitation super-resolution convolutional neural network (SE-SRCNN). The labels above the rectangles of ``Conv2D'' display kernel size, number of filters, and stride size; for instance, ``k9n64s1'' means that the kernel size is $9 \times 9$, the number of filters 64, and the stride size $1 \times 1$. The image height \RevisionA{$h$} and width \RevisionA{$w$} are 64. \RevisionA{The SE block is kept even when the number of inputs $c$ is 1.}}
  \label{fig:model_architecture}
\end{figure}

Our SR model consists of four parts as shown in Fig. \ref{fig:model_architecture}: feature extractors, SE block, nonlinear mapper, and reconstructor. The inputs are an LR temperature $X_0$ and other auxiliary fields \RevisionA{$X_1, \ldots, X_{c-1}$}. The output is an HR temperature field $\hat{Y}$. As a data preprocessing, the bicubic interpolation is applied to all LR inputs to align the size to HR images of \RevisionA{$h \times w \times 1$}, where \RevisionA{$h = w = 64$}. \RevisionA{The total number of inputs is $c$, and our model is called $c$-channel SE-SRCNN. The case of $c=1$ is referred to as the image SR. Even when $c=1$, the SE block is kept. The case of ${c} > 1$ is referred to as the physics SR because relationships among physical quantities are implicitly learned from training data of micrometeorological simulations, as discussed in Section \ref{sec:results_discussion}.}

Each feature extractor transforms the input $X_i$ into a feature $F_i$ (\RevisionA{$i = 0, \ldots, c-1$}), comprising a convolutional layer and a Rectified Linear Unit (ReLU) activation \citep{Glorot2011}:
\begin{equation}
    F_i = \max \left\{0, W^{\rm (FE)}_i \ast X_i + B^{\rm (FE)}_i \right\}, \label{eq:each_features}
\end{equation}
where $\ast$ represents two-dimensional convolution, $W^{\rm (FE)}_i$ is a weight set, and $B^{\rm (FE)}_i$ is a bias set. Note that $W^{\rm (FE)}_i$ and $B^{\rm (FE)}_i$ are different for each feature extractor. The number of convolutional filters is 64 and their kernel size is $9 \times 9$. All features are concatenated to $F$:
\begin{equation}
    F = \left[ F_0, \ldots, F_{c-1} \right], \label{eq:concat_all_features}
\end{equation}
which is \RevisionA{$h \times w \times d$} dimensional (\RevisionA{$d = 64 c$}). \RevisionA{Throughout the paper, multi-dimensional arrays are shown in uppercase, one-dimensional arrays in bold and lowercase, and zero-dimensional arrays (i.e., scalar variables) in lowercase.}

The SE block calculates the dynamic weights for $F$, namely attention weights, whose values are between 0 and 1. These weights are regarded as the \RevisionA{$d$-dimensional} vector $\bm{w}_a$. The calculation method of $\bm{w}_a$ is explained in the next subsection. The features are re-scaled with $\bm{w}_a$:
\begin{equation}
    F_{w_a} = \bm{w}_a \circ F, \label{eq:rescaling_features_by_attentions}
\end{equation}
where $\circ$ stands for depth-wise multiplication. The dimension of $F_{w_a}$ remains unchanged, that is,  \RevisionA{$h \times w \times d$}.

The nonlinear mapper transforms $F_{w_a}$ into a feature $G$. This map is a point-wise ReLU after the convolution,
\begin{equation}
    G = \max \left\{0, W^{\rm (NM)} \ast F_{w_a} + B^{\rm (NM)} \right\}, 
\end{equation}
where $W^{\rm (NM)}$ is a weight set and $B^{\rm (NM)}$ is a bias set. This convolution employs 32 filters with $1 \times 1$ point-wise kernels. The dimension of $G$ is \RevisionA{$h \times w \times 32$}.

The last reconstructor calculates a residual $R$ from $G$ and then the output $\hat{Y}$ by adding the input LR temperature $X_0$ to $R$:
\begin{align}
    R &= W^{\rm (RC)} \ast G + B^{\rm (RC)}, \\
    \hat{Y} &= R + X_0, \label{eq:skip_connection}
\end{align}
where $W^{\rm (RC)}$ is a weight set and $B^{\rm (RC)}$ is a bias set. Single $5 \times 5$ filter is used for the convolution. The dimension of $R$ is the same as that of the input LR temperature $X_0$, i.e., \RevisionA{$h \times w \times 1$}, and hence $\hat{Y}$ also has the same dimension. Note that $R$ can be interpreted as the residual difference between $\hat{Y}$ and $X_0$.

\subsection{Calculation of attention weights} \label{subsec:channel_attention}

The attention weights $\bm{w}_a$ are computed by the SE block \citep{Hu2018}. Its input is the feature $F$, namely the \RevisionA{$h \times w \times d$} dimensional array, where $F_{ijk}$ represents each component.

First, the global average pooling is applied to obtain the statistics $s_k$ for each depth $k$:
\RevisionA{
\begin{equation}
    s_k = \frac{1}{hw} \sum_{i,j} F_{ijk}. \label{eq:statistics_feature_F}
\end{equation}
}
To take into account inter-dependencies between depths, the ReLU function is applied to \RevisionA{$\bm{s} = \left( s_1, \ldots, s_d \right)$}:
\begin{equation}
    \bm{t} = \max \left\{0, W^{({\rm SE}_0)} \bm{s} + \bm{b}^{({\rm SE}_0)} \right\}, \label{eq:transformed_statistics_feature_F}
\end{equation}
where $W^{({\rm SE}_0)}$ is a weight matrix and $\bm{b}^{({\rm SE}_0)}$ is a bias vector. The dimension of $\bm{t}$ remains the same as that of ${\bm s}$. Finally, the sigmoid function is applied to normalize the output $\bm{w}_a$ from 0 to 1:
\begin{equation}
    \bm{w}_a = {\rm sigmoid}\left(W^{({\rm SE}_1)} \bm{t} + \bm{b}^{({\rm SE}_1)} \right), \label{eq:attention_weight}
\end{equation}
where $W^{({\rm SE}_1)}$ is a weight matrix, $\bm{b}^{({\rm SE}_1)}$ is a bias vector, and ${\rm sigmoid}(x) = 1 / [1 + \exp(-x)]$. The output $\bm{w}_a$ is the \RevisionA{$d$-dimensional} vector.

\section{Data preparation and super-resolution model training} \label{sec:training_data_preparation}

The SE-SRCNN was trained and evaluated by using the results of urban micrometeorological simulations. We performed simulations in two different cities: Tokyo and Osaka in Japan. The Tokyo simulations were used for training and evaluating the SE-SRCNN, while the Osaka ones were only for the evaluation to confirm the generalization capability. Section \ref{subsec:urban_meteorological_simulation} briefly describes these simulations. Section \ref{subsec:model_training_evaluation} gives the training method.

\subsection{Urban micrometeorological simulations} \label{subsec:urban_meteorological_simulation}

We employed a multi-scale atmosphere-ocean coupled model named the Multi-Scale Simulator for the Geoenvironment (MSSG) \citep{Takahashi2013, Onishi2012, Sasaki2016, Matsuda2018}. MSSG covers global, meso-, and urban scales. For urban scales, the atmospheric component of MSSG (i.e., MSSG-A) can be used as a building-resolving large-eddy simulation (LES) model coupled with a three-dimensional radiative transfer model \citep{Matsuda2018}. The governing equations for MSSG-A are the conservation equations of mass, momentum, and energy for compressible flows, and the transport equations for mixing ratios of water substances including water vapor, liquid, and ice cloud particles. We performed 85 LESs for Tokyo and 20 LESs for Osaka. A detailed description of the numerical parameters can be found in precursor studies \citep{Onishi2019, Matsuda2018}. 

\begin{figure}[htbp]
  \centering
   \includegraphics[width=14cm]{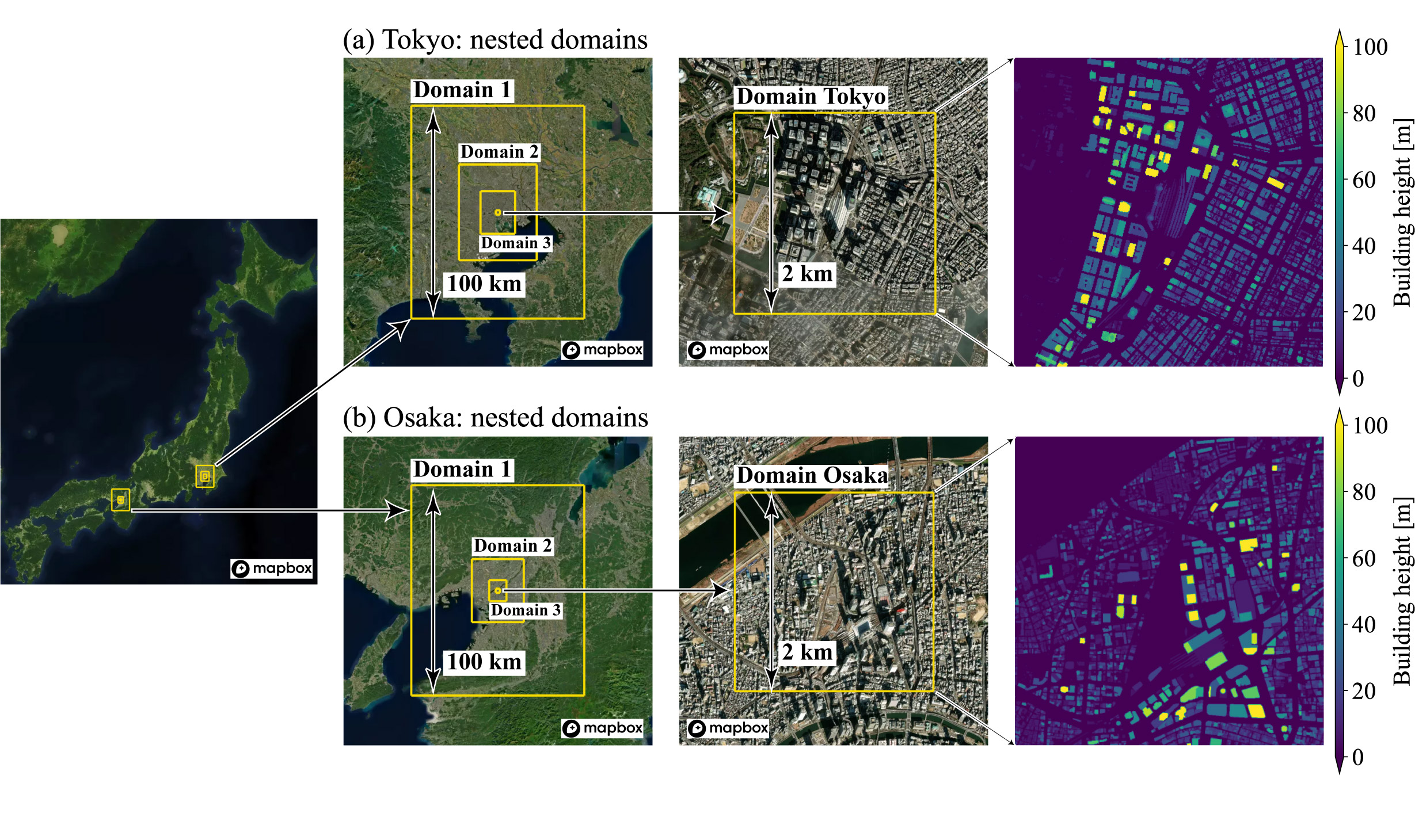}
  \caption{Computational domains for (a) Tokyo and (b) Osaka. The satellite images were obtained from MapBox \citep{mapbox} and OpenStreetMap \citep{openstreetmap}.}
  \label{fig:computational_domain}
\end{figure}

The computational domain in Tokyo was centered at 35.680882$^{\circ}$N and 139.767019$^{\circ}$E (Domain Tokyo in Fig \ref{fig:computational_domain}a), which covered a 2 km $\times$ 2 km horizontal area with the 5-m resolution and the height of 1,500 m with 151 stretched grid points. This domain was embedded in the nested mesoscale simulation domains. Specifically, the mesoscale simulations adopted three two-way-coupled nested systems as shown in Fig. \ref{fig:computational_domain}a: Domain 1 was spanned by horizontal grid points of $100 \times 100$ with the 1-km resolution, Domain 2 $150 \times 150$ with the 300-m resolution, and Domain 3 $200 \times 200$ with the 100-m resolution. For all three nested systems, 55 vertical layers were used for the 40-km height domain, where the lowest layer was located at the 75-m height. The boundary and initial conditions for the mesoscale simulations were taken from the Japan Meteorological Agency (JMA) mesoscale analysis data (MANAL) \citep{jma-mamal}. \RevisionA{The boundary and initial conditions for the LESs were taken from the mesoscale simulations of Domain 3, and the friction drag at the wall surfaces was imposed by the logarithmic law with specified roughness parameters. Data assimilation was not performed in all simulations.} To focus on heat mitigation, 85 hot summer hours were chosen with the maximum hourly temperature that exceeded 35$^\circ$C in the years of 2013-2015. Each LES was run for each targeted hour. The results from the first 10 min were discarded, and the rest 50-min results were analyzed and used to obtain 1-min-average values. That is, each LES produced 50 sets of 1-min-average data.

\begin{figure}[htbp]
  \centering
   \includegraphics[width=9cm]{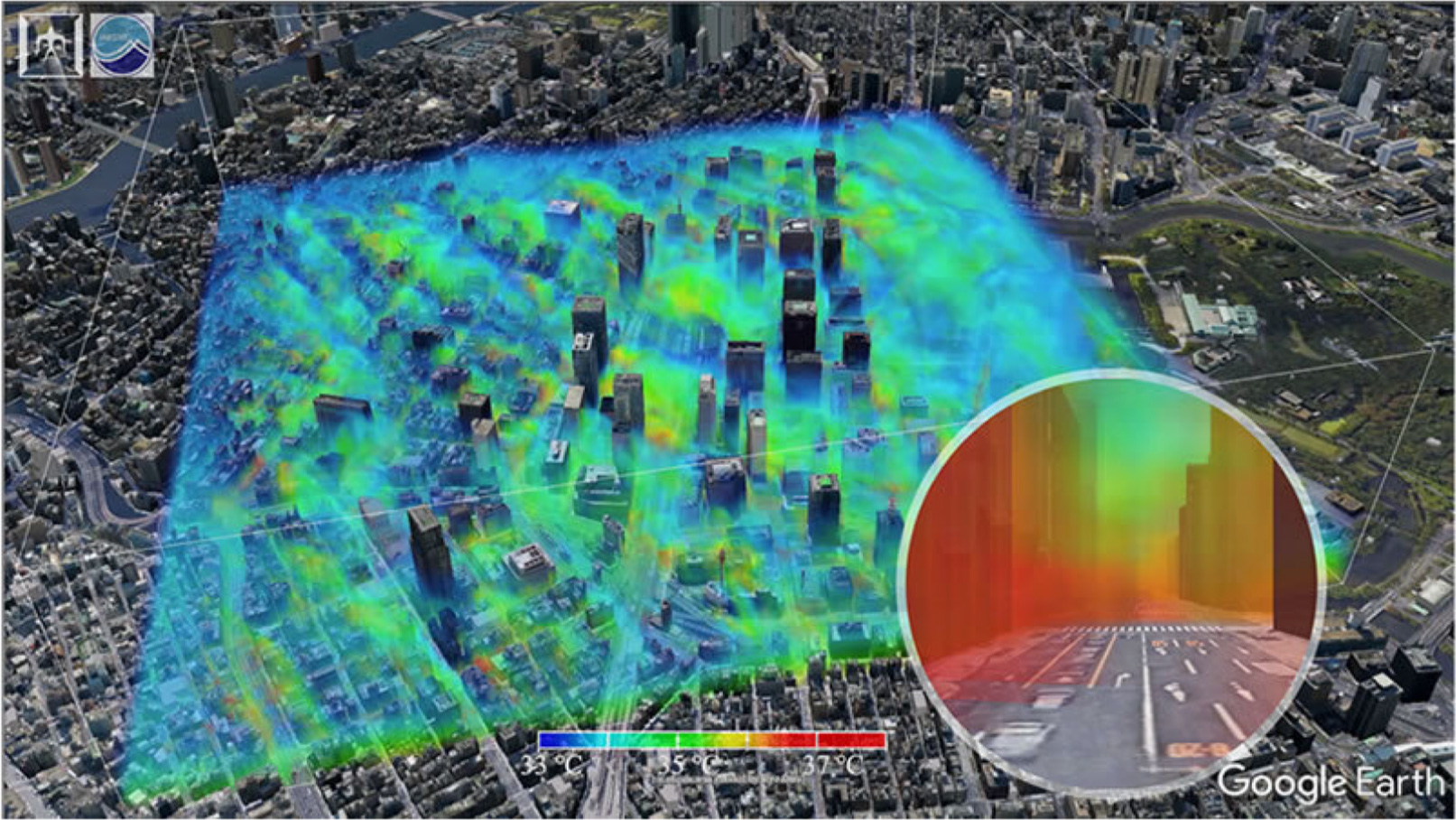}
  \caption{An example of the three-dimensional distribution of the instantaneous air temperature from a high-resolution simulation in the Tokyo domain.}
  \label{fig:tokyo_3d_temperature_image}
\end{figure}

Fig. \ref{fig:tokyo_3d_temperature_image} shows an example of the three-dimensional distribution of the instantaneous air temperature from an HR urban simulation in the Tokyo domain. The Volume Data Visualizer for Google Earth (VDVGE) \citep{Kawahara2015} was utilized for the visualization. The air is warmed up by the land surface, which is heated by solar radiation. The buoyant motion and the turbulent transportation of the warmed air form the puffy structure of the temperature distribution. A simulation movie is also available on YouTube \citep{miraikan-channel}. 

The computational domain in Osaka was centered at 34.70586$^{\circ}$N and 135.4929$^{\circ}$E (Domain Osaka in Fig \ref{fig:computational_domain}b), where the grid configuration was the same as in Tokyo. This Osaka domain was also embedded in the nested three two-way-coupled mesoscale models as shown in Fig. \ref{fig:computational_domain}b: Domain 1 to 3 were spanned by horizontal grid points of $100 \times 100$ with the resolution of 1 km (Domain 1), 300 m (Domain 2), and 100 m (Domain 3). For all three nested systems, 35 vertical layers were used for the 30-km height domain, where the lowest layer was located at the 29-m height. \RevisionA{The boundary and initial conditions for the mesoscale simulations and LESs were given in the same way as for the Tokyo simulations.} Using the same criterion as in Tokyo, 20 hot summer hours were chosen for Osaka in the years of 2017-2018. \RevisionA{Since this period is more recent than the period of Tokyo 2013-2015, the Tokyo and Osaka datasets are almost completely independent of each other.}

\RevisionB{We briefly discuss the complexity of our micrometeorological simulations. Important factors for the complexity are the energy input processes and their space-time scales. In our LES model \citep{Takahashi2013, Onishi2012, Sasaki2016, Matsuda2018}, the energy is input by various processes: the diabatic heating of the three-dimensional radiative transfer, the diabatic heating/cooling associated with the phase transition of water, and the three-dimensional turbulence generation near the building walls. These energy inputs span a variety of scales: streets and buildings having various widths or heights [$O(10-100)$ m], daily solar cycle (24 hours), large-scale flows given by mesoscale simulations through the boundary and initial conditions [$O(1)$ km and $O(1-24)$ hours]. Note that the Reynolds number is about $6 \times 10^{6}$ when the typical length scale is set to 10 m (e.g., street width) and the velocity scale to 10 m s${}^{-1}$. We can conclude that flows simulated by our LES model are much more complex than canonical flows such as channel turbulence.}

\subsection{Training method of super-resolution model} \label{subsec:model_training_evaluation}

The SE-SRCNN was trained and evaluated with sets of the input LR/HR images and the ground truth HR temperature. All HR images were generated from the high-resolution LESs described in the previous subsection. All LR images were obtained by spatially averaging the HR images, where the window size was $4 \times 4$ with the same-size stride, that is, the windows did not overlap each other. \RevisionA{The resolution of LR data is 4 times lower than that of HR data.} The computational cost for a simulation with 4 times coarser spatial grids is roughly 256 ($= 4^4$) times smaller than that for the ground truth HR simulation, which would make a real-time simulation feasible \citep{Onishi2019}. \RevisionB{Note that temperature fields may not have the scale invariance characteristic in urban street scales. The urban streets have rigid scales relating to the street width and the building width/height [$O(10-100)$ m]. This is another reason why we chose the 5-m resolution for HR and the 20-m resolution for LR.}

All LR images were preprocessed as follows. First, the size of LR images was adjusted to the size of HR images $400 \times 400$ by bicubic interpolation. Second, the central $320 \times 320$ area was cropped to exclude influences of the lateral damping layer in the building-resolving LES model. Third, each image was divided into 25 non-overlapping patches of size $64 \times 64$ (\RevisionA{$h = w = 64$}). \RevisionA{These subdivided images were treated as independent.} Finally, each pixel value was normalized from 0 to 1. All HR images were preprocessed in the same way as LR images, except that the bicubic interpolation, namely the first step, was not applied. In our previous study \citep{Onishi2019}, we had taken the central $256 \times 256$ image without partitioning as input, whereas we used here the divided $64 \times 64$ input. This change in the image size reduces the GPU memory consumption and speeds up the model training and inference. We confirmed a slight improvement in the model accuracy compared to the case with $256 \times 256$ inputs.

The inferences of SE-SRCNN are compared by varying combinations of physical quantity inputs: building height distribution, downward shortwave radiation on the surface, temperature field at the 2-m height, and eastward and northward velocity components at the 10-m height. \RevisionA{The building height would help the SR model to recognize the location where wind stagnation leads to temperature rise. Using the downward shortwave radiation, the neural network would detect the sunny area where the surface heats up the air. The wind strength would be useful for estimating the location of turbulence, where strong winds mix the air and cause a drop in temperature. That is, the additional inputs are expected to implicitly provide some physical information to the SE-SRCNN.} 

\RevisionA{The building height field is static and the others are dynamic with respect to time.} The building height and downward shortwave radiation are in high-resolution, while the others are in low-resolution. Even if the downward shortwave radiation is in high-resolution, it can be calculated in a short time with the three-dimensional radiative transfer model \citep{Matsuda2018}.

The SE-SRCNN was trained with Adam optimizer \citep{Kingma2015}, in which the default parameters were used except for $\epsilon = 10^{-7}$ instead of $10^{-8}$. The loss function was set to be the mean squared error between the estimated temperature $\hat{Y}^{(n)}$ and the ground truth $Y^{(n)}$:
\begin{equation}
    \frac{1}{2b} \sum_{n=1}^{b} \left\| \hat{Y}^{(n)} - Y^{(n)} \right\|^2, \label{eq:mse}
\end{equation}
where \RevisionA{$b$} denotes the mini-batch size (\RevisionA{$b=64$}) and $n$ is the index in the mini-batch. All model biases were initialized with zero and all weights were initialized randomly \citep{Glorot2010}. To examine the influences of this random initialization, we trained each SE-SRCNN with five different random seeds\footnote{549736, 480740, 951942, 814099, and 215118. These values were randomly selected.}.

All the Tokyo data were first sorted in chronological order, with the first 60\% as training data, the next 20\% as validation, and the last 20\% as test. \RevisionA{The training data consist of the 51 LESs in the years of 2013-2014, the validation data the 17 LESs in 2014-2015, and the test data the 17 LESs in 2015. Most test data are more than three days in the future from the validation data, and are considered to be statistically independent due to the fact that the time scale of weather is one day \citep{Holton2012}.} In each epoch, 640 samples \RevisionA{of the $64 \times 64$} size were randomly selected from the training and validation data, respectively. In the training step, 10 iterations were performed with a batch size of 64. In the validation step, the loss function was calculated from the 640 samples of the validation data. The model training was stopped by early stopping \citep{Goodfellow2016} with the patience parameter of 300 \RevisionA{epochs}. We confirmed that the results presented below were insensitive to the patience parameter and similar results were obtained with the patience of 200 \RevisionA{epochs}. In all cases, the validation errors were nearly the same as the training ones, suggesting the validity of the hyper-parameter values and neural network architecture. \RevisionA{Note that the choice of hyper-parameter values described above was based on the previous studies \citep{Dong2014, Hu2018, Onishi2019}.} We built the SE-SRCNN with TensorFlow 2.5.0 \citep{tensorflow2015-whitepaper} with the single precision (float32) and performed the training on an NVIDIA A100 40GB PCIe GPU board. \RevisionA{Each model training was completed within three hours.}

\section{Results and discussion} \label{sec:results_discussion}

In this section, we analyze the SE-SRCNN inferences from the test data of the Tokyo LESs and all data of Osaka. None of them were used for training and they all belong to a more recent time interval than the train and validation data. This is to avoid the so-called data leakage. Section \ref{subsec:model_evaluation} performs a comparative analysis of the model performance with different input configurations. \RevisionA{Section \ref{subsec:influence_BH_analysis} investigates influences of building height on the super-resolution. Section \ref{subsec:other_types_LR_datasets} performs the super-resolution of other types of LR data.} Section \ref{subsec:ablation_study} compares the SE-SRCNNs without either the skip connection or the SE block. Section \ref{subsec:discussion_attention} discusses a relationship between the attention weights and the input images.

\subsection{Super-resolution model evaluation} \label{subsec:model_evaluation}

\begin{figure}[htbp]
  \centering
  \includegraphics[width=14cm]{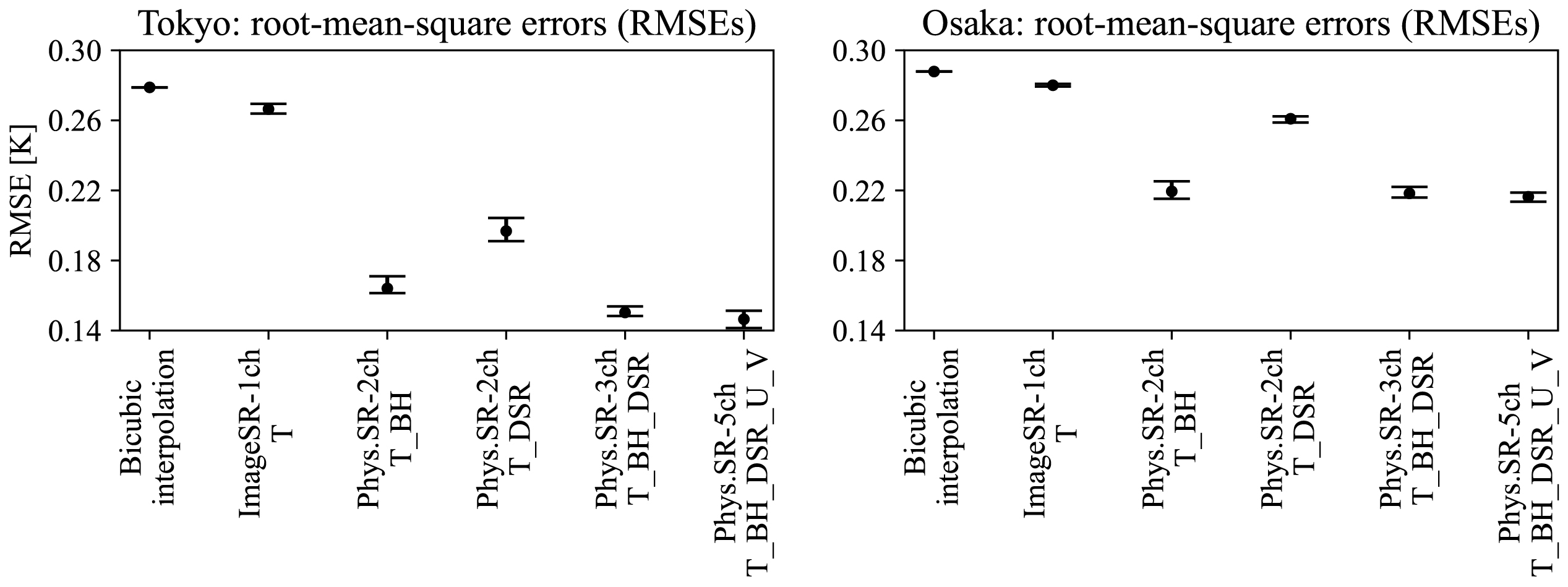}
  \caption{Comparison of root-mean-square errors (RMSEs) in the near-surface temperature for Tokyo and Osaka. The dots represent the RMSEs averaged over the five different random seeds used in the initialization, and the error bars represent the maximum and minimum of these five RMSEs. In the lateral data labels, the symbol ``T'' denotes LR (low-resolution) temperature, ``BH'' HR (high-resolution) building height, ``DSR'' HR downward shortwave radiation, and ``U'' (``V'') LR eastward (northward) velocity component. For instance, the label of ``Phys.SR-2ch T\_BH'' stands for the physics SE-SRCNN with the two inputs of LR temperature and HR building height.}
  \label{fig:comparison_model_rmses}
\end{figure}

Fig. \ref{fig:comparison_model_rmses} shows a comparison of the SE-SRCNN performance across the different configurations of input for Tokyo and Osaka. The dots represent the root-mean-square errors (RMSEs) averaged over the five different random seeds used in the initialization, and the error bars represent the maximum and minimum of these five RMSEs. Table \ref{table:comparison_model_rmses} gives the RMSE values in Fig. \ref{fig:comparison_model_rmses}. As a baseline, the RMSE of bicubic interpolation is added. The magnitude of the error bars is smaller than 10\% of the mean RMSE. \RevisionA{We confirmed that the structural similarity \citep{Wang2004}, which takes into account a spatial correlation between the ground truth and super-resolved temperatures, varied by only about 1\% among the random seeds. Both results suggest that the inference is not severely sensitive to a random seed used in the initialization.} According to the classification of SR methods \citep{Onishi2019}, the SE-SRCNNs in Fig. \ref{fig:comparison_model_rmses} and Table \ref{table:comparison_model_rmses} can be grouped into two: image SR models (SR-lv0), which take only the LR temperature as input, and physics SR models (SR-lv1), which take not only the LR temperature but also other physical quantities as input. For both Tokyo and Osaka, all SR models show smaller RMSEs than the bicubic interpolation, being consistent with our earlier findings \citep{Onishi2019}. Furthermore, all physics SR have smaller RMSEs than the image SR. This result confirms that the dependence of temperature on the other physical quantities is useful for inferring the HR temperature and the physics SR is successful in learning the dependency.

\begin{table}[htbp]
    \centering
    \begin{tabular}{lrrrrrrr}
        \hline
        {} & \multicolumn{4}{l}{Tokyo: RMSE [K]} & \multicolumn{3}{l}{Osaka: RMSE [K]} \\
        {} &          mean &    min &    max &     &     mean &    min &    max \\
        Model                    &               &  &      &        &               &        &        \\
        \hline
        \hline
        Bicubic interpolation    &         0.279 &  0.279 &  0.279 &    &     0.288 &  0.288 &  0.288 \\
        ImageSR-1ch T            &         0.266 &  0.264 &  0.269 &     &    0.280 &  0.279 &  0.281 \\
        Phys.SR-2ch T\_BH         &         0.164 &  0.161 &  0.171 &     &    0.219 &  0.215 &  0.225 \\
        Phys.SR-2ch T\_DSR        &         0.197 &  0.191 &  0.204 &      &   0.261 &  0.259 &  0.262 \\
        Phys.SR-3ch T\_BH\_DSR     &         0.150 &  0.148 &  0.154 &      &   0.218 &  0.216 &  0.222 \\
        Phys.SR-5ch T\_BH\_DSR\_U\_V &         0.146 &  0.141 &  0.151 &     &    0.216 &  0.214 &  0.219 \\
        \hline
    \end{tabular}
    \caption{Comparison of root-mean-square error (RMSE) values shown in Fig. \ref{fig:comparison_model_rmses}. The columns of mean, min, and max represent the mean, minimum, and maximum of RMSEs over the five different random seeds used in the initialization, respectively. See the caption of Fig. \ref{fig:comparison_model_rmses} for the detailed meaning of the row labels.}
    \label{table:comparison_model_rmses}
\end{table}

Fig. \ref{fig:comparison_model_rmses} and Table \ref{table:comparison_model_rmses} indicate that the HR building height is more valuable than the HR downward shortwave radiation for reconstructing the HR temperature. The SE-SRCNN with the LR temperature and HR building height shows the mean RMSEs of 0.164 and 0.219 K for Tokyo and Osaka, respectively (Phys.SR-2ch T\_BH), whereas that with the LR temperature and HR downward shortwave radiation has 0.197 and 0.261 K (Phys.SR-2ch T\_DSR). As for the inference results of Tokyo, we find a slight decrease in the RMSE by taking more physical quantities as input in addition to the building height. When the downward shortwave radiation is added, the RMSE drops from 0.164 to 0.150 K (Phys.SR-3ch T\_BH\_DSR). Moreover, adding the eastward and northward components of velocity, the RMSE decreases from 0.150 to 0.146 K (Phys.SR-5ch T\_BH\_DSR\_U\_V). In contrast, these decreases in RMSE are not clear for Osaka, suggesting that putting more physical quantities into input does not necessarily improve the model accuracy. Winds from the ocean are expected to cool the surrounding city during the day. The bay is located in the south-east of Tokyo (Fig. \ref{fig:computational_domain}a), while that is in the west of Osaka  (Fig. \ref{fig:computational_domain}b). The information outside the input image such as topography may characterize the importance of physical quantities. \RevisionB{In other words, the Tokyo and Osaka data do not share exactly the same statistical features, that is, the ergodicity \citep{Guastoni2021} is not rigorously satisfied. The assessment of data similarity is important to determine the SR model applicability to unknown data; however, this assessment is beyond the scope of the present paper.} 

Other studies have reported that HR inputs facilitate super-resolution. Vandal et al. \citep{Vandal2017} performed the super-resolution of synoptic-scale precipitation with a model stacking SRCNNs. They argued that the HR data of topography elevation made the model inference more precise. Auxiliary HR inputs are employed in the super-resolution of remote sensing data \citep{Yang2017, Bartkowiak2019, Ebrahimy2019, Li2019, Xu2020, Yao2020, Nomura2021}. Nomura and Oki \citep{Nomura2021} carried out the super-resolution of normalized difference vegetation index from satellite images using an SRCNN. \RevisionA{They succeeded in increasing the resolution by a factor of 25 by adding another HR satellite observation to their SRCNN.} Our discussion above is consistent with those results.

\begin{figure}[htbp]
  \centering
  \includegraphics[width=14cm]{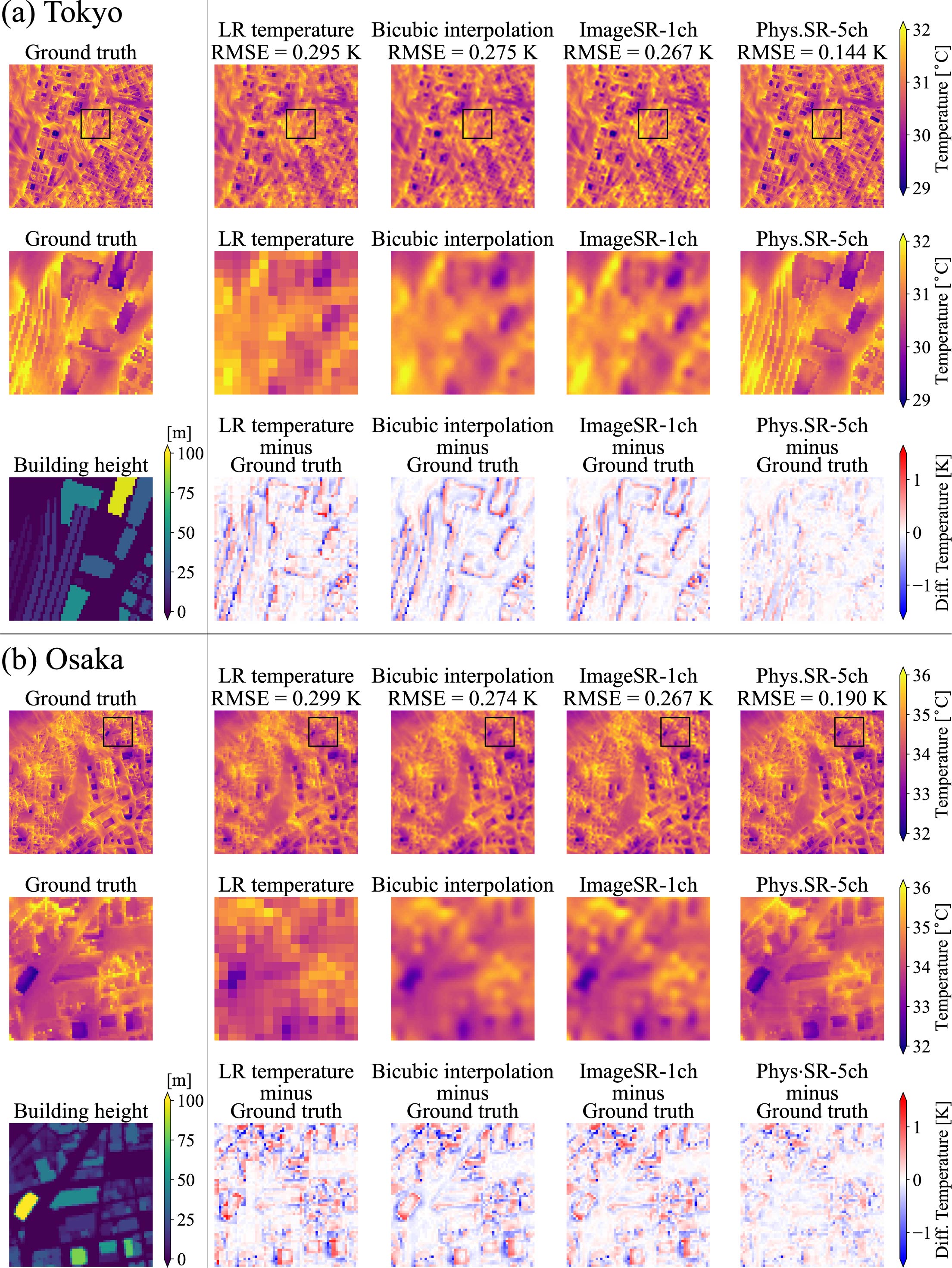}
  \caption{An example of comparison between the ground truth, the LR (low-resolution) temperature, and the inferences by the bicubic interpolation as well as the image and physics SE-SRCNNs: (a) Tokyo and (b) Osaka. The images in the second and third rows of each sub-figure focus on the rectangular areas in the first row. The image size in the first row is $320 \times 320$, which is the whole domain after excluding the region of the lateral damping layer. The image size in the second and third rows is $64 \times 64$.}
  \label{fig:comparison_sr_images}
\end{figure}

Fig. \ref{fig:comparison_sr_images} shows a comparison between the ground truth, the LR temperature, and the inferences by the bicubic interpolation as well as the image and physics SR. The 5-channel SE-SRCNN is used here as the physics SR. In Fig. \ref{fig:comparison_sr_images}a and b, the images in the second row are created by enlarging the rectangular areas in the first row, and the third-row images show the errors within these rectangles, together with the building height distributions. As for Tokyo (Fig. \ref{fig:comparison_sr_images}a), the physics SR greatly reduces the errors near the boundaries of buildings, compared to the bicubic interpolation and the image SR. Although this error decrease is not much clear for Osaka (Fig. \ref{fig:comparison_sr_images}b), the physics SR still improves temperature in the vicinity of buildings. This result suggests that the physics SE-SRCNN utilizes the distribution of building height to selectively reduce the errors of temperature near the building walls. 

\begin{figure}[htbp]
  \centering
  \includegraphics[width=6.5cm]{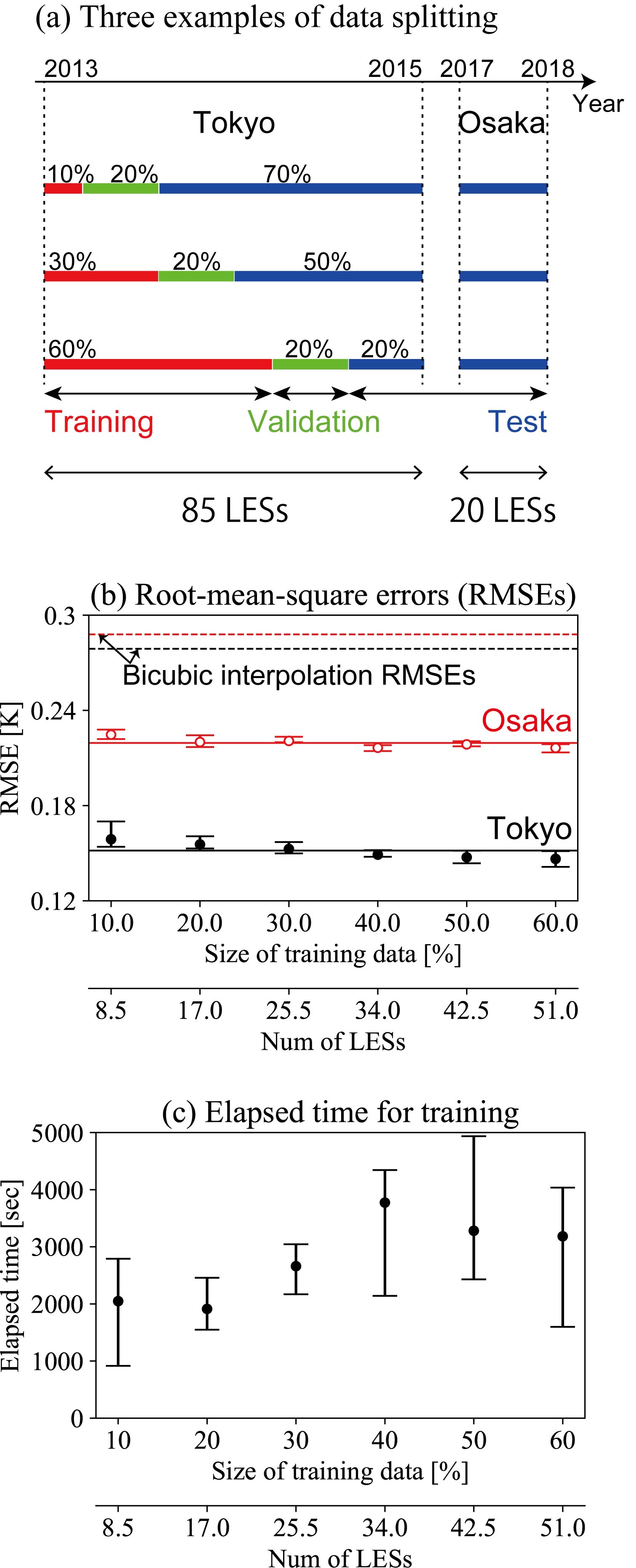}
  \caption{\RevisionB{(a) Three examples of the data partitioning. Each partition is used for one iteration of the repeated holdout method \citep{Cerqueira2020, Tashman2000}. The partitioning with the 60\% training data (bottom) was used for the evaluation in Fig. \ref{fig:comparison_model_rmses} and Table \ref{table:comparison_model_rmses}. (b) Root-mean-square errors (RMSEs) and (c) elapsed time for training against various data partitions. The error bars in (b) and (c) represent the maximum and minimum values over the five random seeds used during the initialization. The dashed lines in (b) show the RMSEs of bicubic interpolation, while the solid ones are the RMSE averages; for both kinds of lines, the red (black) ones represent the Osaka (Tokyo) values. The averaged RMSEs give estimations of the generalization error, according to the repeated holdout method \citep{Cerqueira2020, Tashman2000}.}}
  \label{fig:rmse_with_various_data_size}
\end{figure}

\RevisionB{We investigate RMSEs with various sizes of the training data by using the 5-channel SE-SRCNN, leading to an estimation of the generalization error as discussed below. Fig. \ref{fig:rmse_with_various_data_size}a shows three examples of data partitioning, where the Tokyo data are split into training, validation, and test data with preserving the chronological order. The bottom example, namely the partition with the 60\% training data, was used in the above evaluation (Fig. \ref{fig:comparison_model_rmses} and Table \ref{table:comparison_model_rmses}). Fig. \ref{fig:rmse_with_various_data_size}b shows the RMSEs against the various training data sizes. The same training method as in Section \ref{subsec:model_training_evaluation} was used for each data splitting. The RMSE is not strongly dependent on the data size, and all RMSEs of Tokyo and Osaka are sufficiently smaller than the bicubic interpolation values (dashed lines). The result indicates that the 60\% training data are sufficient to estimate the test RMSEs. Fig. \ref{fig:rmse_with_various_data_size}c shows the elapsed time for each training. As the training data size increases, the training time tends to increase, though it depends on the random seed. Depending on the initial weights generated from the random seed, the number of epochs until the learning saturates is different, resulting in a difference in the elapsed time determined by the early stopping.}

\RevisionB{The generalization error can be estimated using various data partitions with sample mean (e.g., cross validation). Since future data in time series generally depend on the past ones, careful treatments are needed to split time series. Cerqueira et al. \citep{Cerqueira2020} demonstrated that the repeated holdout method \citep{Tashman2000} may give more accurate estimation of the generalization error than the cross validation. One iteration of the repeated holdout is the estimation of test scores using a data partition as in Fig. \ref{fig:rmse_with_various_data_size}a, and the generalization error is evaluated as the average over all iterations. The solid lines in Fig. \ref{fig:rmse_with_various_data_size}b represent the averaged RMSEs. We confirm that the averages for Tokyo and Osaka are nearly equal to the RMSEs using the 60\% training data, implying that the generalization errors are not much different from these RMSEs.}

Even a cutting-edge building-resolving LES such as of Matsuda et al. \citep{Matsuda2018} has an inherent error of 0.23 K in a heat index when the 5-m spatial resolution is used. Liu et al. \citep{Liu2012} compared air temperatures simulated by the LES model in high- and low-resolution with observed ones and reported that the RMSE decreased by about 0.2 K when the higher resolution was used. These results indicate that a required precision in the SR downscaling would be around 0.2 K. Our 5-channel physics SR shows the mean RMSEs close to this criterion: 0.146 K for Tokyo and 0.216 K for Osaka (Table \ref{table:comparison_model_rmses}).

We briefly discuss the computation time of the 5-channel SR. It takes $6.0 \times 10^{-4}$ sec to make a prediction of $64 \times 64$ size and $4.3 \times 10^{-3}$ sec for the data preprocessing described in Section \ref{subsec:model_training_evaluation}. Thus, the total elapsed time to infer a $64 \times 64$ HR temperature is about $5.0 \times 10^{-3}$ sec. For instance, a 30-min simulation makes 30 sets of 1-min-average data with the size of $320 \times 320$. The SR inference totally takes about 3.75 sec to process these data. \RevisionA{It takes about 2.6 min to make 30-min LR prediction data, which are the input for our SR model. The total computation time is less than 3 min; hence, the combination of LR simulations with the 5-channel SE-SRCNN would make a real-time operation feasible.} See Onishi et al. \citep{Onishi2019} for a detailed discussion. \RevisionA{Uncertainty quantification for inference \citep{Maulik2020, Sun2020} is important for real-world applications. Maulik et al. \citep{Maulik2020} presented a mixture density network and predicted the parameters of the probability density without sampling, which saved the inference time. It would be important to estimate the computation time including uncertainty quantification, using such an efficient model.}

\subsection{Influences of building height on super-resolution} \label{subsec:influence_BH_analysis}

\RevisionA{In this subsection, we investigate influences of building height on improving the accuracy of super-resolution. To eliminate the effects of the other physical quantities, the 2-channel SE-SRCNN is employed, which has the two inputs: LR temperature and HR building height.}

\RevisionA{An influence of binarization is first examined. The building height (BH) at each location is binarized as follows:
\begin{equation}
    \begin{cases}
        1 & (BH > 0), \\
        0 & (\text{otherwise}).
    \end{cases} \label{eq:binarizedBH}
\end{equation}
The 2-channel SE-SRCNN was trained and tested using the binarized BH, where the training method was the same as in Section \ref{subsec:model_training_evaluation}. Table \ref{table:result_binarizedBH} compares the RMSEs with and without the BH binarization. The mean RMSE becomes slightly larger for Tokyo, from 0.164 to 0.180K, which is still much smaller than the bicubic interpolation value of 0.279 K. For Osaka, the mean RMSE increases from 0.219 to 0.294 K, which is even larger than the bicubic interpolation of 0.288 K. The difference between the maximum and minimum of RMSEs also increases from 0.010 to 0.028 K, meaning that the inference becomes sensitive to a random seed used in the weight initialization.}

\begin{table}[htbp]
    \centering
    \scalebox{0.9}[0.9]{
    \begin{tabular}{lrrrrrrr}
    \hline
    {} & \multicolumn{4}{l}{Tokyo: RMSE [K]} & \multicolumn{3}{l}{Osaka: RMSE [K]} \\
    {} &            mean &    min &    max &      &      mean &    min &    max \\
    Model                        &                 &     &   &        &                 &        &        \\
    \hline
    Bicubic Interpolation        &           0.279 &  0.279 &  0.279 &      &     0.288 &  0.288 &  0.288 \\
    Phys.SR-2ch T\_BH            &           0.164 &  0.161 &  0.171 &      &     0.219 &  0.215 &  0.225 \\
    Phys.SR-2ch T\_BH (binarized) &           0.180 &  0.177 &  0.182 &      &     0.294 &  0.276 &  0.304 \\
    \hline
    \end{tabular}
    }
    \caption{\RevisionA{Comparison of the RMSEs with and without the BH (building height) binarization. The first and second rows, Bicubic interpolation and Phys.SR-2ch T\_BH, are the same as in Table \ref{table:comparison_model_rmses}. The last row shows the RMSEs with the BH binarization. See the caption of Table \ref{table:comparison_model_rmses} for the meaning of the column labels.}}
    \label{table:result_binarizedBH}
\end{table}

\begin{figure}[htbp]
  \centering
  \includegraphics[width=12cm]{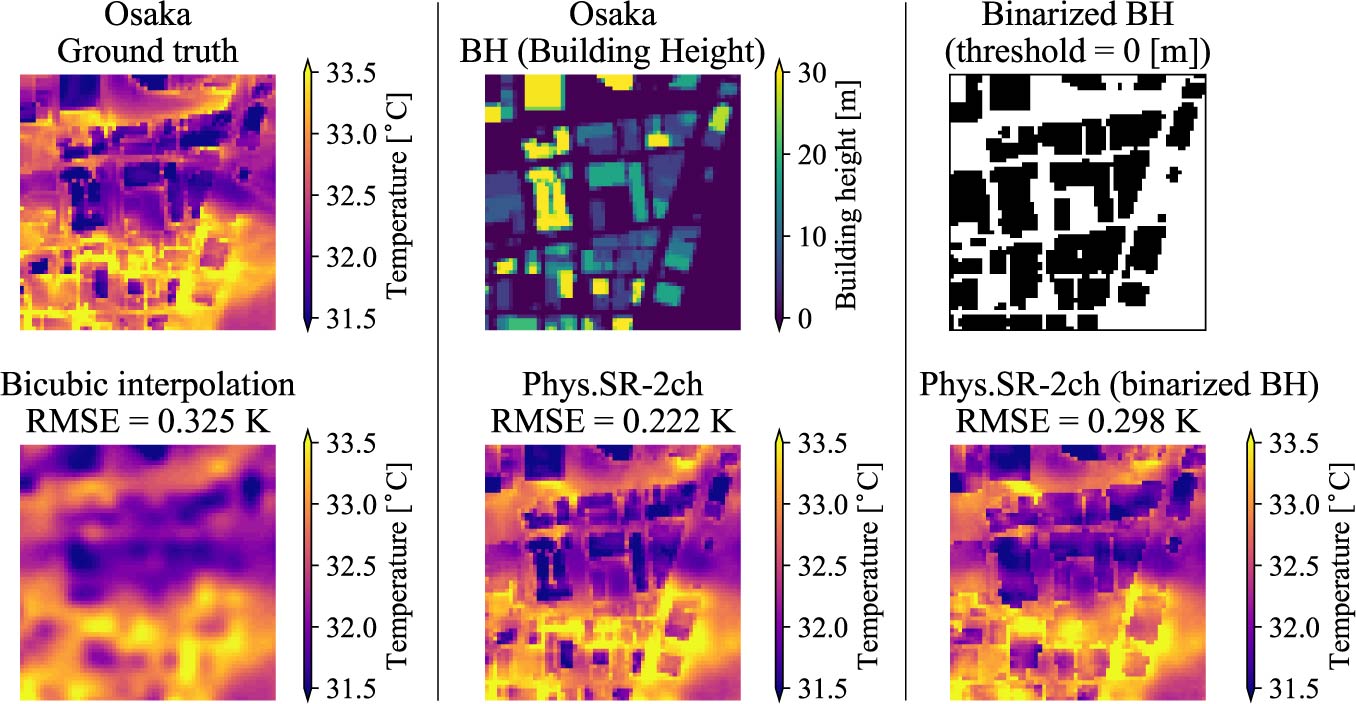}
  \caption{\RevisionA{An example of super-resolved temperatures for Osaka with and without the BH (building height) binarization: (left) the ground truth and bicubic interpolation temperatures, (middle) the building height and inferred temperature, and (left) the building height and inferred temperature with the BH binarization. All images are $64\times64$.}
  \label{fig:example_SR_of_binarizedBH}}
\end{figure}

\RevisionA{Fig. \ref{fig:example_SR_of_binarizedBH} shows an example of super-resolved temperatures for Osaka with and without the BH binarization. When the BH is not binarized, the SE-SRCNN successfully reconstructs the HR temperature field. Even with the binarization, the inferred temperature reflects the small-scale structures of the BH; its RMSE of 0.298 K is larger than 0.222 K for the non-binarized case and close to the bicubic interpolation of 0.325 K. This deterioration in accuracy is likely attributed to the fact that the binarized BH does not correctly describe the building boundaries. For instance, multiple buildings connected by a low-height corridor can be considered as one building, depending on the threshold in Eq. (\ref{eq:binarizedBH}), which was set to zero. In other words, the SR model inference is sensitive to the threshold of the binarization. This sensitivity is not observed for Tokyo because there are few structures between the 0 and 10-m height and the binarized field does not strongly depend on the threshold. The RMSE for Tokyo is small regardless of the BH binarization, implying that a BH value itself does not have a direct influence on the super-resolution. Rather, a trained model may utilize a BH field to determine the appropriate boundary for each building, as discussed below.}

\begin{figure}[htbp]
  \centering
  \includegraphics[width=14cm]{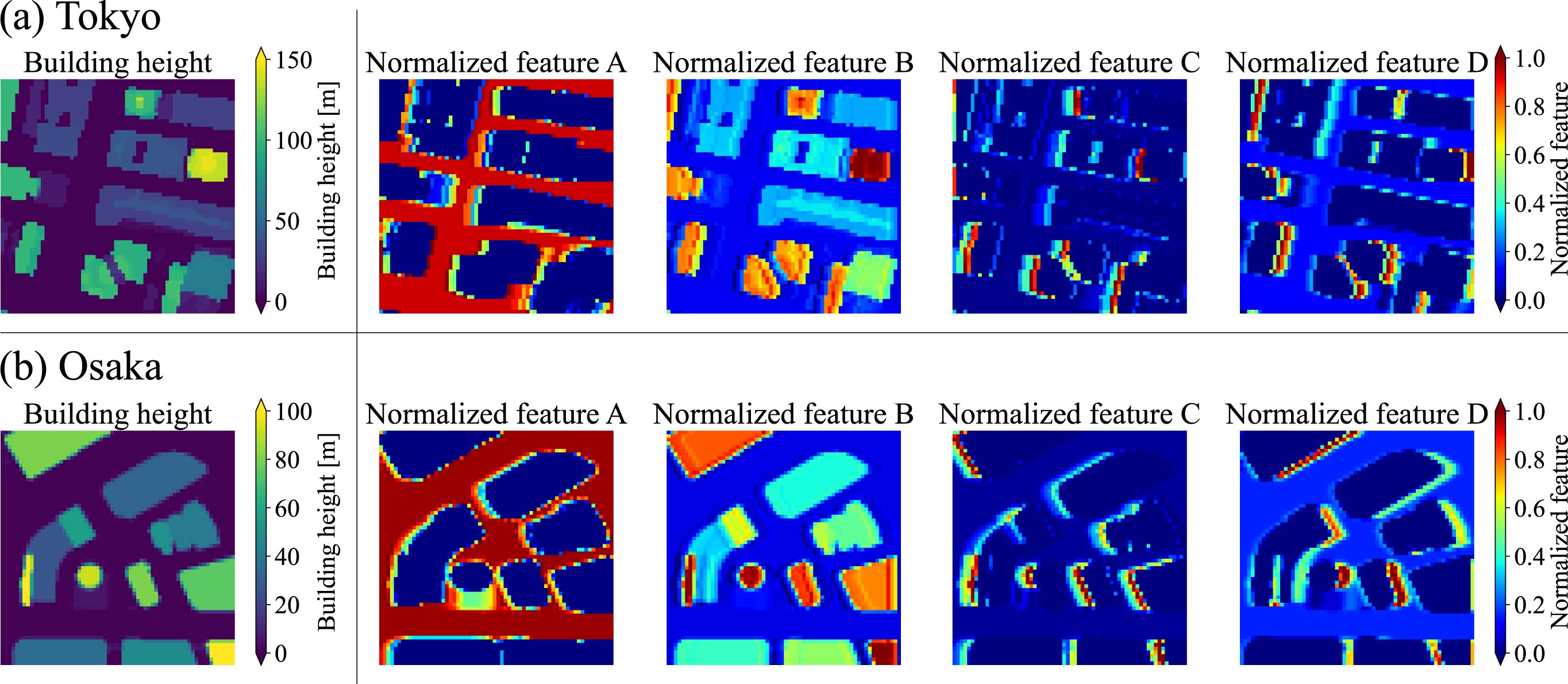}
  \caption{\RevisionA{An example of BH (building height) feature maps: (a) Tokyo and (b) Osaka areas of $64\times64$. Feature map means an intermediate output from the BH feature extractor in Eq. (\ref{eq:each_features}). Four of the 64 features are selected above. Each feature map is normalized from 0 to 1. The input BH is also shown in the most left-hand side of (a) and (b).}}
  \label{fig:feature_maps_BH}
\end{figure}

\RevisionA{Generally, a well-trained CNN extracts useful features from its input \citep{Zeiler2014}; for instance, some kernels obtain texture information from relatively smooth signals, whereas some detect edges from nearly discontinuous signals. Fig. \ref{fig:feature_maps_BH} shows the intermediate outputs from the BH feature extractor in Eq. (\ref{eq:each_features}), where four of the 64 feature maps are selected. Feature A appears to respond strongly to the ground level and Feature B reflects the BH values. Feature C and D describe the building edges, and even the curved edges in Osaka are successfully detected (Fig. \ref{fig:feature_maps_BH}b). These edge detectors automatically determine the appropriate boundary for each building, resulting in the accurate reconstruction of HR temperature fields. If we binarized building height as in Eq. (\ref{eq:binarizedBH}), which determines the edges, it might be inappropriate for inference and lead to poor accuracy (Table \ref{table:result_binarizedBH} and Fig. \ref{fig:example_SR_of_binarizedBH}).}

\subsection{Super-resolution for other types of low-resolution data} \label{subsec:other_types_LR_datasets}

\RevisionA{The accuracy of SR models may depend on methods to create LR data \citep{Fukami2019, Wang2020, Kim2021}. To check sensitivities on types of LR data, we apply the 5-channel SE-SRCNN to subsampled data, in addition to locally averaged data having lower-resolution than in the previous subsections. The local average was the same operation as described in Section \ref{subsec:model_training_evaluation}, i.e., the spatial average of HR images, where the window size was $r \times r$ with the same-size stride. An HR image of $64\times64$ is then coarse-grained into the LR one of $64/r \times 64/r$. The value of $r$ was set not only to 4 but also to 8. Subsampled data were made by sampling the HR images with intervals of $r \times r$. The SE-SRCNN was trained and tested for each LR dataset using the same training method as in Section \ref{subsec:model_training_evaluation}.}

\RevisionA{Table \ref{table:model_scores_various_LR_data} compares the RMSEs for the four LR datasets. The RMSEs of bicubic interpolation are added to the column of ``BI'' as baselines. For the locally averaged LR data, even with $r=8$, the mean RMSEs are much smaller than the baselines: For Tokyo, the mean RMSE is 0.209 K and the baseline is 0.362 K; for Osaka, the mean RMSE is 0.295 K and the baseline is 0.383 K. For the subsampled LR data, the RMSEs are much smaller than the corresponding baselines again; however, they become slightly larger than the ones with the local average. This result is likely due to aliasing errors. The LR images are distorted by aliasing; it is impossible to separate high-frequency from low-frequency signals if the higher component exceeds the Nyquist frequency. Kim et al. \citep{Kim2021} introduced a pixel-loss that incorporates the fact that subsampled LR data are pointwise accurate. Their loss function may help to reduce the errors.}

\begin{table}[htbp]
    \centering
    \scalebox{0.9}[0.9]{
    \begin{tabular}{lccccccccc}
    \hline
    {} & \multicolumn{5}{l}{Tokyo: RMSE [K]} & \multicolumn{4}{l}{Osaka: RMSE [K]} \\
    {} &            mean &    min &    max &              BI &       &     mean &    min &    max &              BI \\
    Low-resolution data             &                 &        &     &   &                   &                 &        &        &                   \\
    \hline
    Local average ($r=4$) &           0.146 &  0.141 &  0.151 &             0.279 &   &        0.216 &  0.214 &  0.219 &             0.288 \\
    Subsampling ($r=4$)   &           0.171 &  0.166 &  0.179 &             0.341 &    &       0.260 &  0.260 &  0.260 &             0.351 \\
    Local average ($r=8$) &           0.209 &  0.205 &  0.215 &             0.362 &     &      0.295 &  0.293 &  0.297 &             0.383 \\
    Subsampling ($r=8$)   &           0.244 &  0.241 &  0.248 &             0.436 &      &     0.358 &  0.355 &  0.360 &             0.463 \\
    \hline
    \end{tabular}
    }
    \caption{\RevisionA{Comparison of the RMSEs for the four LR datasets: the local averaged and subsampled data with $r = 4$ or $8$. The HR temperature of $64\times64$ is reconstructed from the LR one of $64/r\times64/r$. The values of $r = 4$ and $8$ mean the $\times$4 and $\times$8 super-resolution, respectively. The 5-channel SE-SRCNN was trained for each dataset, and the RMSE was calculated using it. The column of ``BI'' represents the RMSEs of bicubic interpolation. For the meaning of the other columns, see the caption of Table \ref{table:comparison_model_rmses}. The values in the first row are the same as the ones in the last raw of Table \ref{table:comparison_model_rmses}.}
    \label{table:model_scores_various_LR_data}}
\end{table}

\begin{figure}[htbp]
  \centering
  \includegraphics[width=14cm]{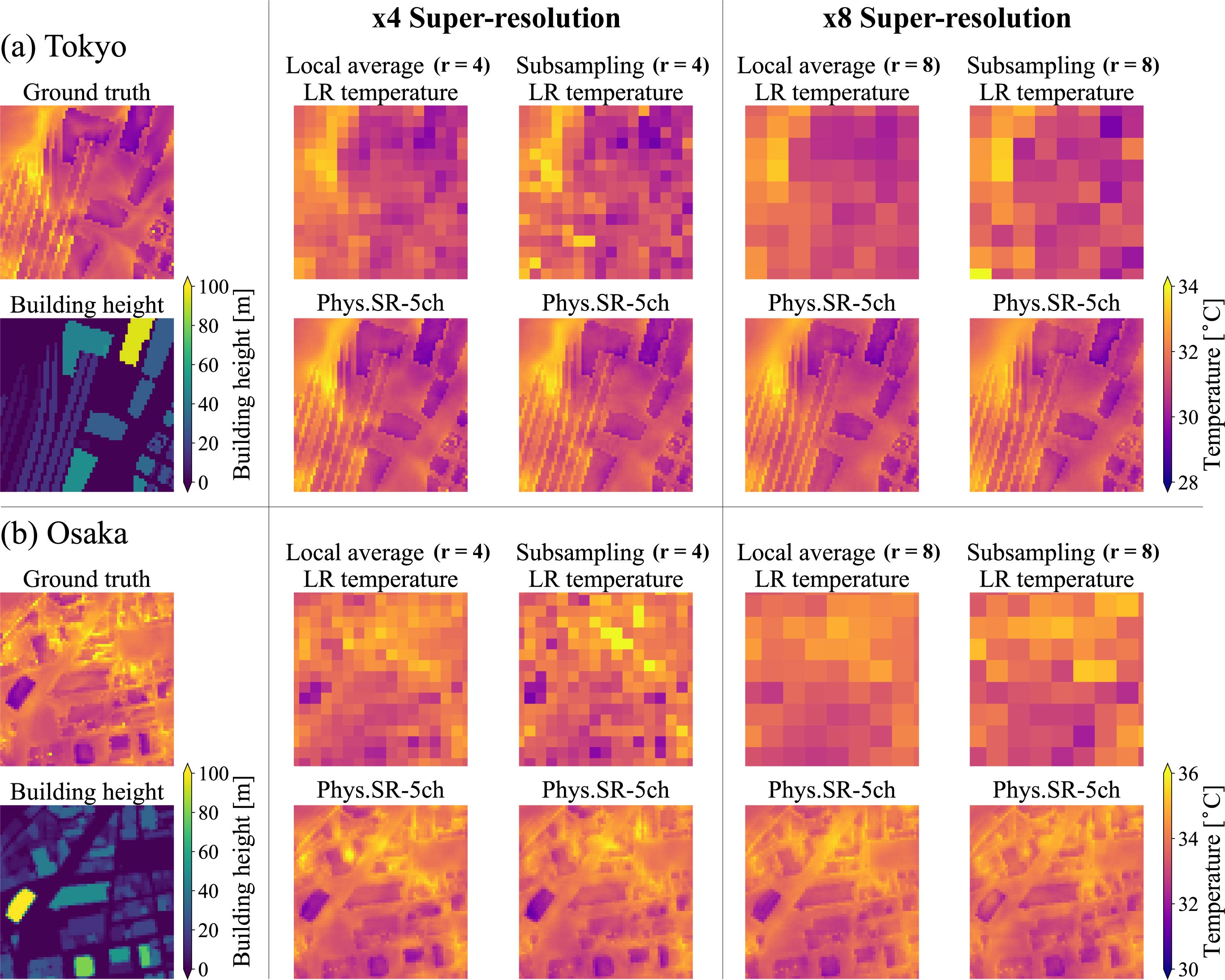}
  \caption{\RevisionA{An example of pairs of low-resolution (LR) and super-resolved (SR) temperatures: (a) Tokyo and (b) Osaka. The 5-channel SE-SRCNN was employed. The SR temperatures are displayed below the corresponding LR temperature inputs. The ground truth temperature and the building height are displayed on the most left side. The four LR datasets are examined: the local averaged and subsampled LR data with the factor $r$ of 4 or 8. The HR temperature of $64\times64$ is reconstructed from the LR one of $64/r\times64/r$.}}
  \label{fig:example_SR_various_LR_datasets}
\end{figure}

\RevisionA{Figs. \ref{fig:example_SR_various_LR_datasets}a and \ref{fig:example_SR_various_LR_datasets}b show an example of pairs of LR and SR temperatures for Tokyo and Osaka, respectively, together with the ground truth and building height fields. Even when $r = 8$, the inferred temperature fields reconstruct the small-scale structures of buildings, which are not observed in the LR temperature inputs.}

\subsection{Ablation study} \label{subsec:ablation_study}

We trained the 5-channel SE-SRCNN without either the skip connection or the SE block to confirm their validity. Without the skip connection, Eq. (\ref{eq:skip_connection}) changes to $Y = R$. Without the SE block, all components of $\bm{w}_a$ in Eq. (\ref{eq:rescaling_features_by_attentions}) become a constant $1$. Table \ref{table:comparison_model_rmses_ablation} shows the RMSEs of these SE-SRCNNs with the original one.

\begin{figure}[htbp]
  \centering
  \includegraphics[width=14cm]{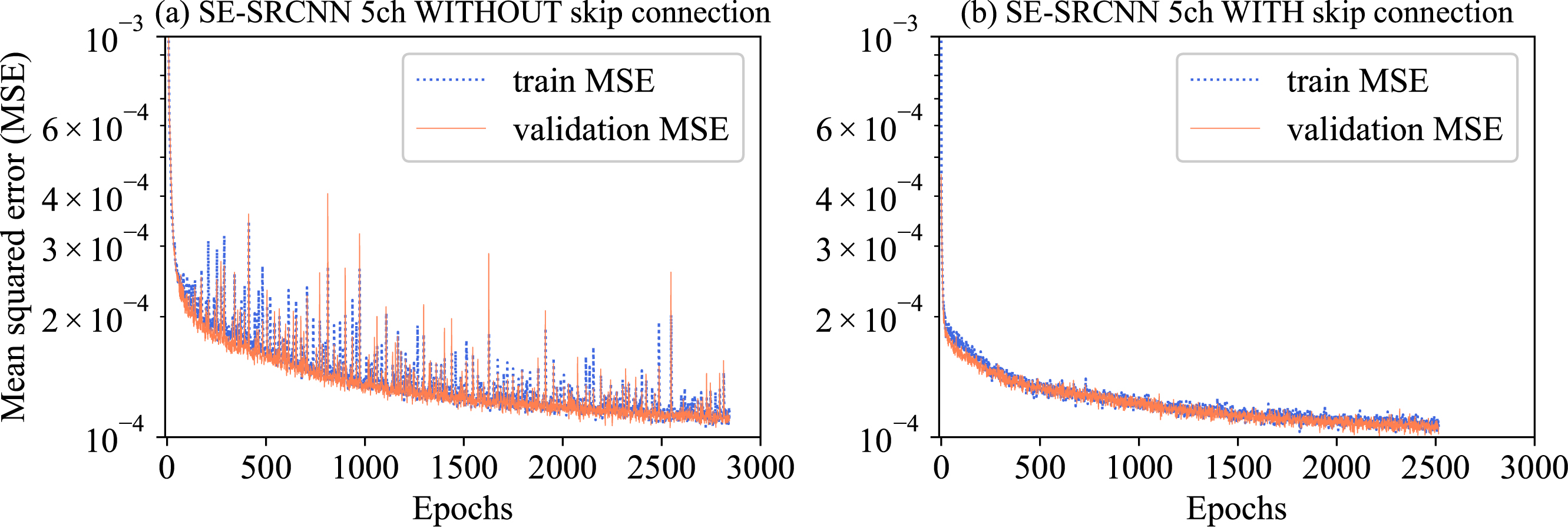}
  \caption{Comparison between the learning curves of the 5-channel SE-SRCNN (a) without and (b) with the skip connection. In each plot, the dotted (solid) line shows the loss values for the train (validation) data. The non-dimensional mean squared error (MSE) was used as the loss function, as in Eq. (\ref{eq:mse}).}
  \label{fig:comparison_learning_curve_ablation}
\end{figure}

The skip connection slightly improves the model accuracy; more importantly, it makes the model learning more stable and faster. Fig. \ref{fig:comparison_learning_curve_ablation} shows the learning curves with and without the skip connection for one of the five random seeds. Similar curves were obtained for the other random seeds. Without the skip connection (Fig. \ref{fig:comparison_learning_curve_ablation}a), the learning curves are more volatile and the model learning is stopped at about 2800 epochs. In contrast, when the skip connection is incorporated (Fig. \ref{fig:comparison_learning_curve_ablation}b), the fluctuations in the learning curves are significantly reduced and the learning is stopped at about 2500 epochs. \RevisionA{To quantify the volatility of the learning curves, we calculated the standard deviation from the 5-point moving average, confirming that the standard deviation decreased by about 10 times when the skip connection was employed.} Drozdzal et al. \citep{Drozdzal2016} showed that more skip connections suppressed the variation in the learning curves and increased in the model convergence speed for a biomedical image segmentation. Li et al. \citep{Li2018} demonstrated that skip connections reduced the irregularity of loss functions for an image classification. All results including ours support that skip connections make the model learning stable and fast.

\begin{table}[btp]
    \centering
    \begin{tabular}{lrrrrrrr}
        \hline
        {} & \multicolumn{4}{l}{Tokyo: RMSE [K]} & \multicolumn{3}{l}{Osaka: RMSE [K]} \\
        {} &          mean &    min &    max &    &      mean &    min &    max \\
        SE-SRCNN-5ch   &               &        &    &    &               &        &        \\
        \hline
        \hline
        Original      &         0.146 &  0.141 &  0.151 &       &  0.216 &  0.214 &  0.219 \\
        No skip connection &         0.149 &  0.147 &  0.150 &   &      0.218 &  0.216 &  0.221 \\
        No SE block   &         0.151 &  0.146 &  0.154 &        & 0.225 &  0.218 &  0.235 \\
        \hline
    \end{tabular}
    \caption{Comparison of root-mean-square errors (RMSEs) in the near-surface temperature across the original 5-channel SE-SRCNN and ones without either the skip connection or the SE (squeeze-and-excitation) block. See the caption of Table \ref{table:comparison_model_rmses} for the detailed meaning of the column labels.}
    \label{table:comparison_model_rmses_ablation}
\end{table}

\begin{figure}[htbp]
  \centering
  \includegraphics[width=14cm]{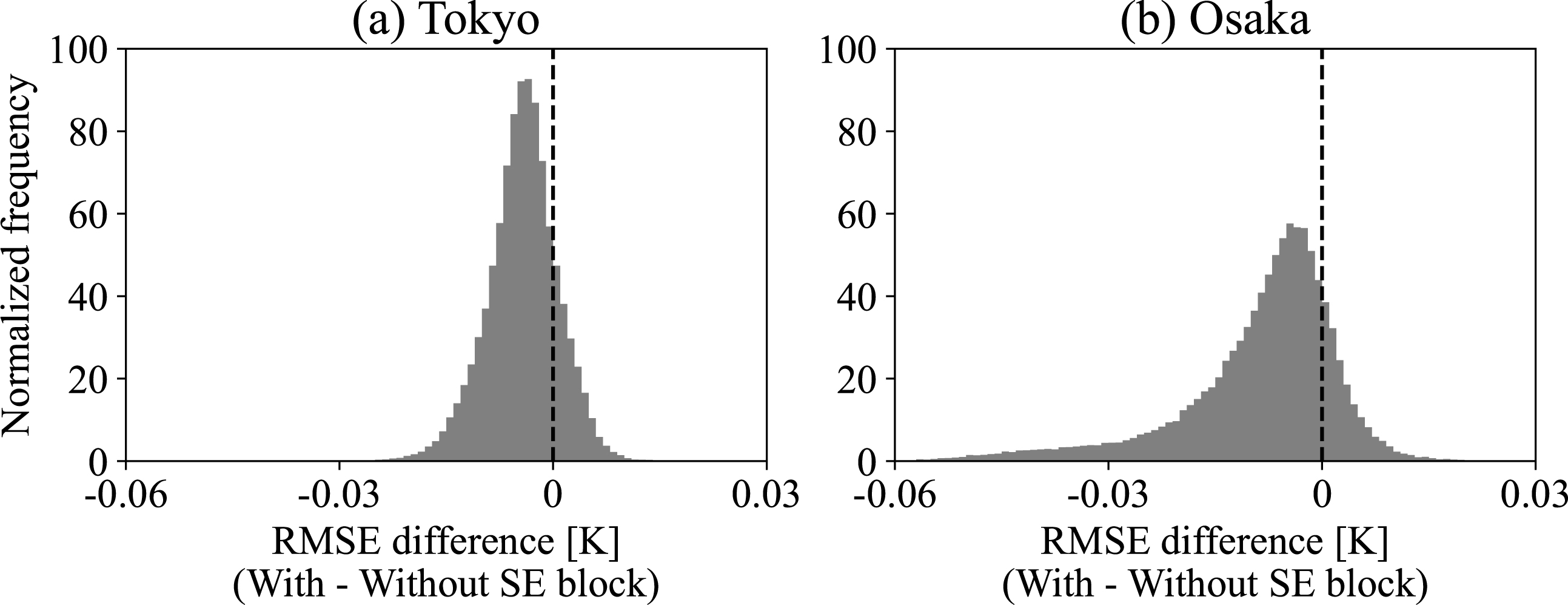}
  \caption{Distributions of difference in the root-mean-square error (RMSE) of the near-surface temperature between the 5-channel SE-SRCNNs with and without the SE block: (a) Tokyo and (b) Osaka. Each distribution was obtained with the SE-SRCNNs of all the five random seeds used in the initialization.}
  \label{fig:srcnn_vs_senet_rmse_diff_comparison}
\end{figure}

The SE block improves the model accuracy: The mean RMSE decreases from 0.151 to 0.146 K for Tokyo and from 0.225 to 0.216 K for Osaka (Table \ref{table:comparison_model_rmses_ablation}). Fig. \ref{fig:srcnn_vs_senet_rmse_diff_comparison} shows the distributions of difference in the RMSE. The distributions were obtained with the SE-SRCNNs of the five random seeds. Each difference value is defined by subtracting an RMSE without the SE block from the one with the SE block. The distributions clearly have negative biases for both Tokyo (Fig. \ref{fig:srcnn_vs_senet_rmse_diff_comparison}a) and Osaka (Fig. \ref{fig:srcnn_vs_senet_rmse_diff_comparison}b). \RevisionA{In fact, the p-values of the binomial test are zero for Tokyo and Osaka with the single precision (float32), suggesting that the medians of both distributions in Fig. \ref{fig:srcnn_vs_senet_rmse_diff_comparison} are significantly negative.} These results mean that the SE block reduces the RMSE and improves the model accuracy. Zhang et al. \citep{Zhang2018RCAN} developed an SR neural network with SE blocks and argued that the channel attention improved the model performance for general images. Our result is consistent with theirs. Furthermore, the distribution for Osaka (Fig. \ref{fig:srcnn_vs_senet_rmse_diff_comparison}b) has the fat tail on the negative side, suggesting that the SE block reduces the number of outliers and makes the model inference more robust. It should be noted that the stability and convergence speed of the model training were not largely affected by the presence or absence of the SE block.

\subsection{Relationship between attention weights and input images} \label{subsec:discussion_attention}

One of the purposes of introducing attention mechanisms is to dynamically adjust model weights on the basis of its input, which is not possible with simple CNNs that use static weights. In this subsection, we examine a relationship between changes in the input images and the weights of the channel attention.

\begin{figure}[htbp]
  \centering
  \includegraphics[width=14cm]{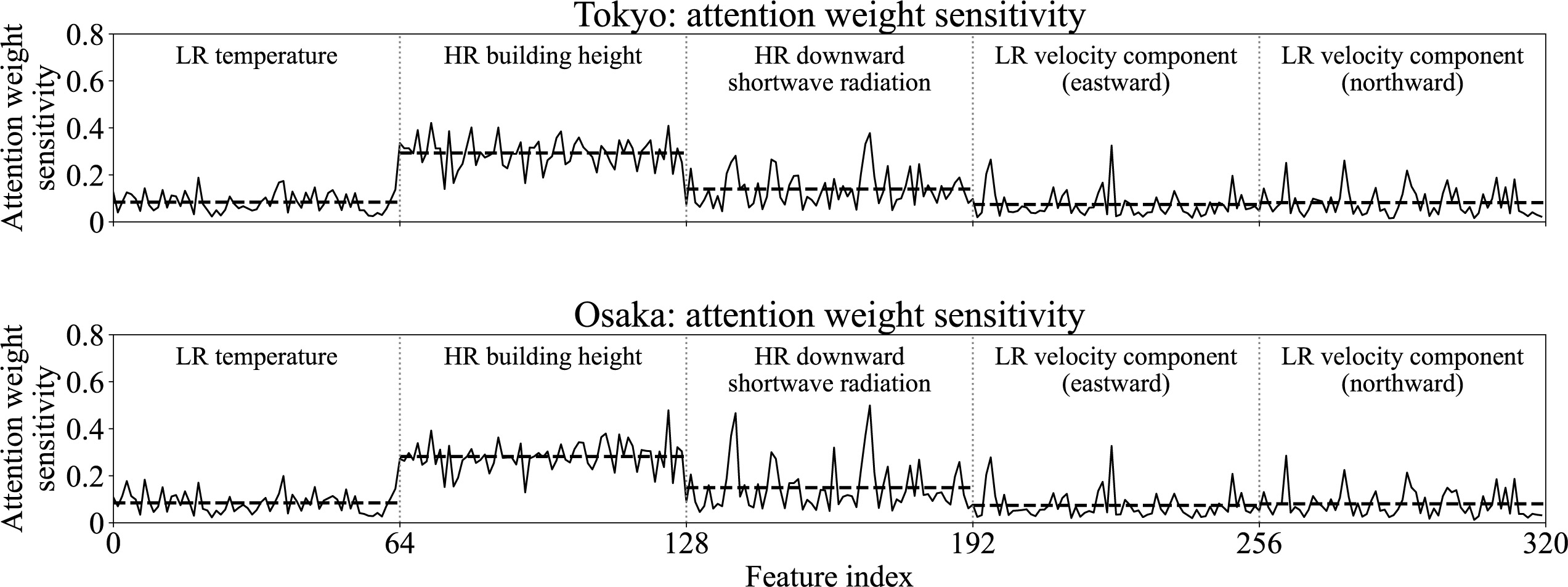}
  \caption{Sensitivity of the attention weights over all features in Eq. (\ref{eq:concat_all_features}). The top panel is for Tokyo and the bottom for Osaka. The sensitivity for each feature is defined as the difference between 5\% and 95\% tiles of the attention weight over all test data. The horizontal dashed lines represent the means of the sensitivity over the 64 features in Eq. (\ref{eq:each_features}) for the corresponding physical quantities.}
  \label{fig:attention_sensitivity}
\end{figure}

Fig. \ref{fig:attention_sensitivity} shows the sensitivity of the attention weights over all features in Eq. (\ref{eq:concat_all_features}), where the sensitivity for each feature is defined as the difference between 5\% and 95\% tiles of the attention weight over all test data. The number of features for each physical quantity is 64 because we employed the 64 convolutional filters in Eq. (\ref{eq:each_features}). The horizontal dashed lines in Fig. \ref{fig:attention_sensitivity} represent the means of the sensitivity over the 64 features for the corresponding physical quantities. On average the attention weight sensitivity of the building height is the largest for both Tokyo and Osaka. In other words, the importance of the building-height features changes the most, depending on the input image set. We further investigate the sensitivity of the building-height attention weights.

\begin{figure}[tbp]
  \centering
  \includegraphics[width=14cm]{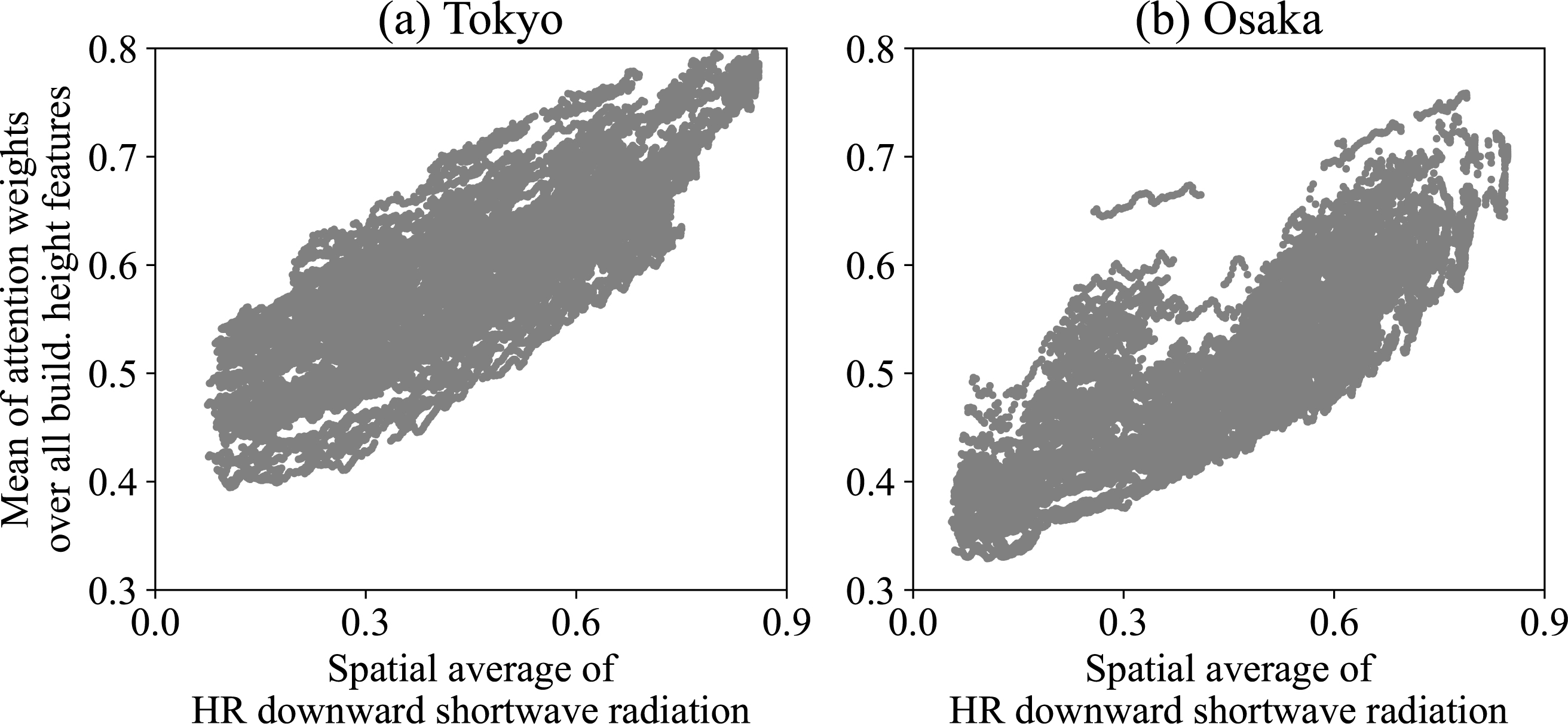}
  \caption{Scatter plots of the mean of the building-height attention weights against the spatial average of the downward shortwave radiation: (a) Tokyo and (b) Osaka. For each input, the mean of the attention weights is calculated by averaging over the 64 weights of the building height.}
  \label{fig:attention_scatter_comparison}
\end{figure}

SE blocks \citep{Hu2018} compute attention weights from spatial averages of features, as in Eqs. (\ref{eq:statistics_feature_F}), (\ref{eq:transformed_statistics_feature_F}), and (\ref{eq:attention_weight}). We expect that the attention weights change depending on the spatial averages of input images. Fig. \ref{fig:attention_scatter_comparison} shows scatter plots of the mean of the attention weights against the spatial average of the downward shortwave radiation, where for each input the mean of the attention weights is given by the average over the 64 weights of the building height. For both Tokyo (Fig. \ref{fig:attention_scatter_comparison}a) and Osaka (Fig. \ref{fig:attention_scatter_comparison}b), the attention weights tend to increase as the downward shortwave radiation becomes larger. In other words, the building-height attention weights are sensitive to changes in the spatial average of the downward shortwave radiation. The building-resolving LES model of this paper calculates the three-dimensional radiative transfer \citep{Matsuda2018}. When the downward shortwave radiation increases, the reflection on the building walls becomes stronger and the contrast between sun and shade may have a greater influence on the temperature distribution. This analysis suggests that the physics SR model learns the dependence of temperature on the building height through the downward shortwave radiation. For the other physical quantities, we did not find a simple relationship against the attention weights of the building height. Nonlinear relationships among multiple physical quantities possibly affect the attention weights; however, examining these relationships is beyond the present study.

\section{Conclusions} \label{sec:conclusions}

We have proposed a super-resolution (SR) model based on a convolutional neural network and applied it to the near-surface temperature from building-resolving large-eddy simulations (LESs). Our physics SR model incorporates a skip connection \citep{He2016}, a channel attention mechanism \citep{Hu2018}, and separated feature extractors for the inputs of temperature, building height, downward shortwave radiation, and horizontal velocity. This SR model was trained only with the LESs for a Tokyo metropolitan area, whereas it was evaluated with the LESs not only for Tokyo but also for Osaka, another metropolitan city in Japan. The physics SR model infers more accurate high-resolution (HR) temperature fields than the bicubic interpolation and the image SR that takes only the temperature as input. Since the Osaka data were not used for the training, this result suggests the generalizability of our SR model. Moreover, the short inference time suggests the potential of the SR model to facilitate a real-time HR prediction in metropolitan cities by combining with a low-resolution (LR) micrometeorological model. \RevisionA{The SR model was found to be more accurate than the bicubic interpolation for both locally averaged and subsampled LR data, not only for the 20- to 5-m super-resolution, but also for the 40- to 5-m.} Except for the temperature input, the building height is the most important factor in reconstructing the HR temperature field. The input of building height can selectively reduce the errors in temperature near the building boundaries. \RevisionA{The analysis using the binarized building height reveals that the trained SR model considers a building boundary from its height information.} The analysis of attention weights indicates that the importance of building height is sensitive to the downward shortwave radiation. The contrast between sun and shade becomes stronger with the increase in solar radiation, which may affect the temperature distribution. \RevisionA{Our SR model does not explicitly include laws of physics into the loss function and is different from physics-informed neural networks in Raissi et al. \citep{Raissi2019}.} The present results suggest that the model inference reflects the physical relationships and our SR model learns some physics from the training data of the LESs, \RevisionA{in which the governing equations were numerically integrated.}

An important future work is the super-resolution of another physical quantity \RevisionA{such as velocity.} In recent years, several studies have proposed artificial neural networks (ANNs) for super-resolving the three-dimensional velocity \citep{Fukami2021, Liu2020, Bode2021, Bai2020, Ferdian2020}. Investigating the applicability of these models to building-resolving LESs will be a first step towards real-time three-dimensional HR predictions. \RevisionB{The super-resolved three-dimensional velocity is necessary to evaluate physics losses, i.e., residuals of the governing equations, and the simultaneous super-resolution of the velocity and temperature will lead to a physics-informed neural network \citep{Raissi2019}.} In the present study, the spatial average \RevisionA{or subsampling} was utilized to derive LR images from the HR ones. This kind of downsampling methods has been used in fluid-related SR research \citep{Fukami2019, Fukami2021, Liu2020, Bode2021, Kim2021, Jiang2020, Xie2018, Ferdian2020, Vandal2017, Leinonen2020, Stengel2020}. Wang et al. \citep{Wang2020} argued that when LR images are created by downsampling, ANNs learn the inverse of downsampling algorithms and hence LR and HR images should be generated from separated numerical simulations. \RevisionA{Kim et al. \citep{Kim2021} developed an unsupervised learning with unpaired HR and LR data, where the LR data were generated from the LESs and the HR ones were from the direct numerical simulations.} From a practical viewpoint, it may be better to create LR train data from micrometeorological simulations with some data assimilation technique that maintains the difference between LR and HR simulations.

\section*{Declaration of competing interest}

The authors declare that they have no known competing financial interests or personal relationships that could have appeared to influence the work reported in this paper.

\section*{Acknowledgements}

This work was supported by the JSPS KAKENHI (grant numbers 20H05751 and 20H02074). The building-resolving urban simulations were performed on the Earth Simulator system of the Japan Agency for Marine-Earth Science and Technology (JAMSTEC).

% \appendix
% \section{Sample Appendix Section}
% \label{sec:sample:appendix}

\bibliographystyle{elsarticle-num} 
\bibliography{references}





% \end{thebibliography}
\end{document}